\documentclass[preprint,12pt]{elsarticle}

\usepackage{hyperref}

\usepackage{amssymb}

\usepackage{amsmath}
\usepackage{subcaption}
 \usepackage{amsthm}

\usepackage{lineno}

\newcommand{\PD}[2]{\frac{\partial #1}{\partial #2}}
\newcommand{\eq}[1]{\begin{linenomath}\begin{equation} #1 \end{equation}\end{linenomath}}

\newcommand{\algn}[1]{\begin{linenomath}\begin{align} #1 \end{align}\end{linenomath}}

\newcommand{\B}[1]{\mathbf{#1}}

\newcommand{\twobytwomatrix}[4]{\left(\begin{array}{cc} #1 & #2 \\ #3 & #4\end{array}\right)}

\newcommand{\arr}{\rightarrow}

\newcommand{\ol}[1]{\overline{#1}}

\newcommand{\half}{\tfrac{1}{2}}
\newcommand{\bs}[1]{\boldsymbol{#1}}
\newcommand{\tl}[1]{\tilde{#1}}
\newcommand{\EXP}[1]{\left\langle #1 \right\rangle}
\newcommand{\pr}{\partial}
\newcommand{\lrb}[1]{\left(#1\right)}
\newcommand{\lrsq}[1]{\left[#1\right]}
\newcommand{\lrc}[1]{\left\{#1\right\}}
\newcommand{\al}{\alpha}
\newcommand{\Half}{\frac{1}{2}}

\newcommand{\eqsys}[1]{\begin{subequations}\algn{#1}\end{subequations}}
\newcommand{\eqsyslab}[2]{\begin{subequations} \label{eq:#2} \algn{#1}\end{subequations}}

\newcommand{\be}{\beta}
\newcommand{\abs}[1]{\left| #1 \right|}

\usepackage{afterpage}
\usepackage{cleveref}
\usepackage{float}

\journal{Ocean Modelling}

\begin{document}

\begin{frontmatter}



\title{On Divergence- and Gradient-Preserving Coarse-Graining for Finite Volume Primitive Equation Ocean Models}


\author[1]{Stuart Patching\corref{cor1}%
}
\ead{s.patching17@imperial.ac.uk}

\cortext[cor1]{Corresponding author}

\affiliation[1]{organization={Department of Mathematics, Imperial College London}, 
            city={London},
            country={United Kingdom}}

\begin{abstract}
We consider the problem of coarse-graining in the context of finite-volume fluid models. If a variable is defined on a high-resolution grid it may be coarse-grained so that it is defined on a grid of lower resolution. In general this will cause some information about the variable to be lost. In particular, horizontal divergences, gradients or other operators calculated on the coarse grid after projecting may differ from those calculated on the fine grid. In some cases we are able to choose averaging weights for coarse-graining such that the coarse-grid operators will give a result approximating that of the corresponding fine-grid operators applied on the fine grid. In this work we derive general conditions on the averaging weights that allow the divergence and gradient to be preserved. These conditions are applied to a regular triangular mesh with B-grid variable placement in which the fine-grid resolution is some integer multiple $N$ of the coarse-grid resolution. For this case we find particular values for the averaging weights that preserve the divergence, and a set of different averaging weights that preserve the gradient. We observe that the vertical component of the curl is also preserved by the same coarse-graining that preserves divergence. These coarse-grainings are applied to data from FESOM2 simulations and we demonstrate that using the particular coarse-grainings derived herein gives an overall reduction in $L^1$ error when compared with other methods. 

\end{abstract}




\end{frontmatter}


\section{Introduction}
\label{sec:intro}
 Coarse-graining is an important tool in modern ocean modelling, as it allows us to compare variables defined on a high-resolution `fine' grid with variables defined on a low-resolution `coarse' grid. This is often necessary in studies of eddy parameterisation \citep[e.g.][]{BERLOFF2005123,berloff2005random,berloff_ryzhov_shevchenko_2021,mana2014toward,cotter2018modelling,cotter2019numerically} as a way of identifying the `eddy forcing' that quantifies the effects of missing scales that are not resolved in low-resolution model runs. For example, the dynamical flow decomposition method \citep{BERLOFF2005123,berloff_ryzhov_shevchenko_2021} uses the results of a high-resolution model run as the reference `truth'; the coarse-grid model run is then corrected towards the coarse-grained truth and the amount by which the solution needs to be corrected gives the eddy forcing; the statistical properties of the eddy forcing can then be used to configure a stochastic parameterisation \citep{berloff2005random}. Since in this application the model run is corrected towards coarse-grained truth variables, it is important that what we use as the truth on the coarse grid retains some properties of the original truth as defined on the fine grid. Another approach to parameterisation is used in \cite{mana2014toward}, in which the eddy forcing is calculated by considering the differences between the advection and diffusion terms when calculated on the coarse-grid from the coarse-grained fields, and the coarse-grained versions of the fine-grid advection and viscosity terms. In that paper the coarse graining is configured so that the coarse-grained fields have the same total integral as the fine-grid field. Further methods such as Stochastic Advection by Lie Transport \citep{holm2015} also use coarse-graining in order to compute the parameters governing the stochastic advection \citep{cotter2018modelling,cotter2019numerically}. The approach there involves looking at the differences between the Lagrangian trajectories computed using fine-grid velocities and those computed using coarse-grained velocities. \\
\par
The present work may also be applicable to studies of passive tracers in cases in which the resolution required by the tracer equation is lower than that required by the ocean model from which the velocity field is obtained. It will be particularly important in studies of biogeochemistry \citep[][]{VICHI200789,ENGLAND_MAIERREIMER2001,England_Rahmstorf} because the incompressibility condition, which ensures mass conservation, determines the vertical velocity in the primitive equation models. Explicitly, the vertical velocity is calculated according to:  
\algn{
    w = - \int_{-H}^{z}\nabla\cdot\B{u} dz' \label{eq:vert_vel}
}
Thus, any errors that are introduced into the divergence will then manifest in the vertical velocity and so cause spurious effects in the vertical transport. Coarse-graining in such a way as to preserve the divergence would provide a method to avoid such errors. \\
\par
There is further potential to apply the present work to ocean-atmosphere coupling. In coupled models the oceanic component is typically run at a higher resolution that the atmosphere component \citep{CABOS,Xue2019}. In order to couple the two models, it is therefore necessary to re-map the appropriate fields onto the corresponding grids \citep{OASIS}. When performing this re-mapping it may be desirable to preserve certain properties such as the curl of the wind stress. In particular, if there is a horizontal wind stress $\bs{\tau}(x,y)$ at the surface, then potential vorticity is generated in the ocean by the curl of this wind stress, $\PD{\tau_2}{x} - \PD{\tau_1}{y}$. It is therefore important to preserve this curl when coupling the two components of the model. While it is true that the atmosphere component would need to be interpolated onto the (higher-resolution) ocean grid rather than coarse-grained, the techniques developed herein could be used to verify that the interpolation preserves the curl, or to impose conditions on the type of interpolation used. Coarse-graining and interpolation also have potential applications in climate studies such as CMIP \citep{CMIP2016} which use large data sets. In some cases it may be impractical to store the entire data output, and so having a method of coarse-graining the data to a lower resolution in such a way that preserves essential properties would be useful in reducing the total amount of data. This may find applications where the full high-resolution field is not required for the relevant application; or, if we are additionally able to develop suitable interpolation procedures which also preserve these properties then we would have a prescription for returning the data to a higher resolution where this is needed. The problem of finding such interpolations from coarse to fine grid that preserve desired properties provides a potential extension of the present work, and we do not explore this particular application in detail here. 
\\
\par
Another quantity of interest in PE models is the pressure gradient $\nabla p$, which acts as a force on the horizontal velocities. For flows in geostrophic balance the velocity is given by $\B{u}_{g} := \frac{1}{f}\hat{\mathbf{k}}\times \nabla p$, which gives a good approximation to the full velocity field \citep[see][]{vallis2017atmospheric}. Thus it is desirable to compute the pressure gradient accurately on the coarse grid in order that spurious effects should not be introduced into the geostrophic velocity. We shall seek to define a coarse-graining of scalars that avoids such errors in the gradient.

\subsection{Problem Statement} \label{sec:defs}
\begin{figure}[H]
    \centering
    \begin{subfigure}{.5\textwidth}
      \centering
        \includegraphics[width=.9\textwidth]{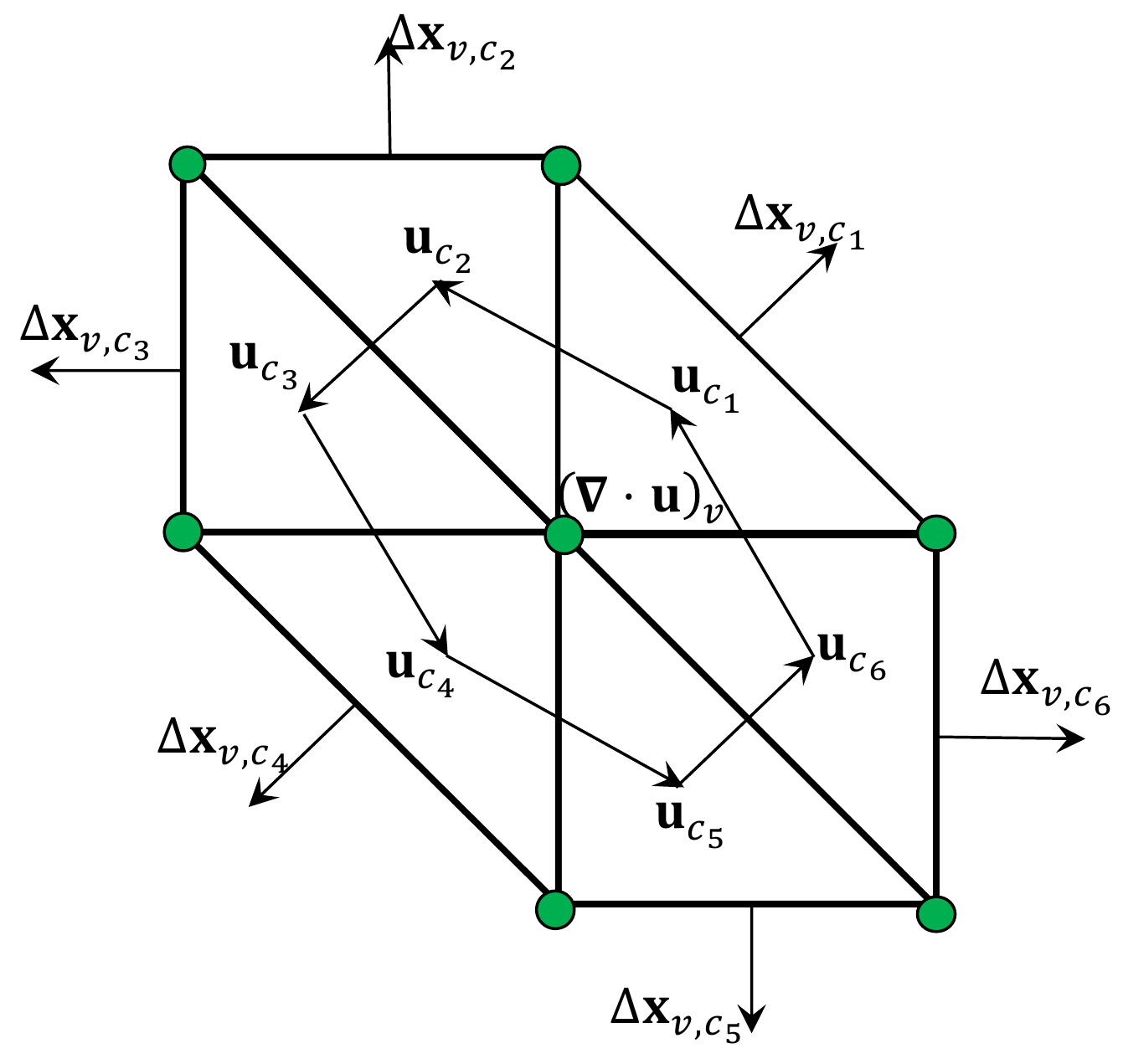}
         \captionsetup{width=0.9\linewidth}
        \caption{Definition of divergence for finite volume methods on Arakawa-B grids. Velocities $\B{u}$ are defined at cell centres, so divergence is defined via an integral over the vertex cell and an application of the divergence theorem gives the result as an integral over the boundary of the vertex cell of the dot product of the velocity with the outward-pointing normal.}
        \label{fig:div_def}
    \end{subfigure}%
       \begin{subfigure}{.5\textwidth}
        \centering
        \includegraphics[width=.9\textwidth]{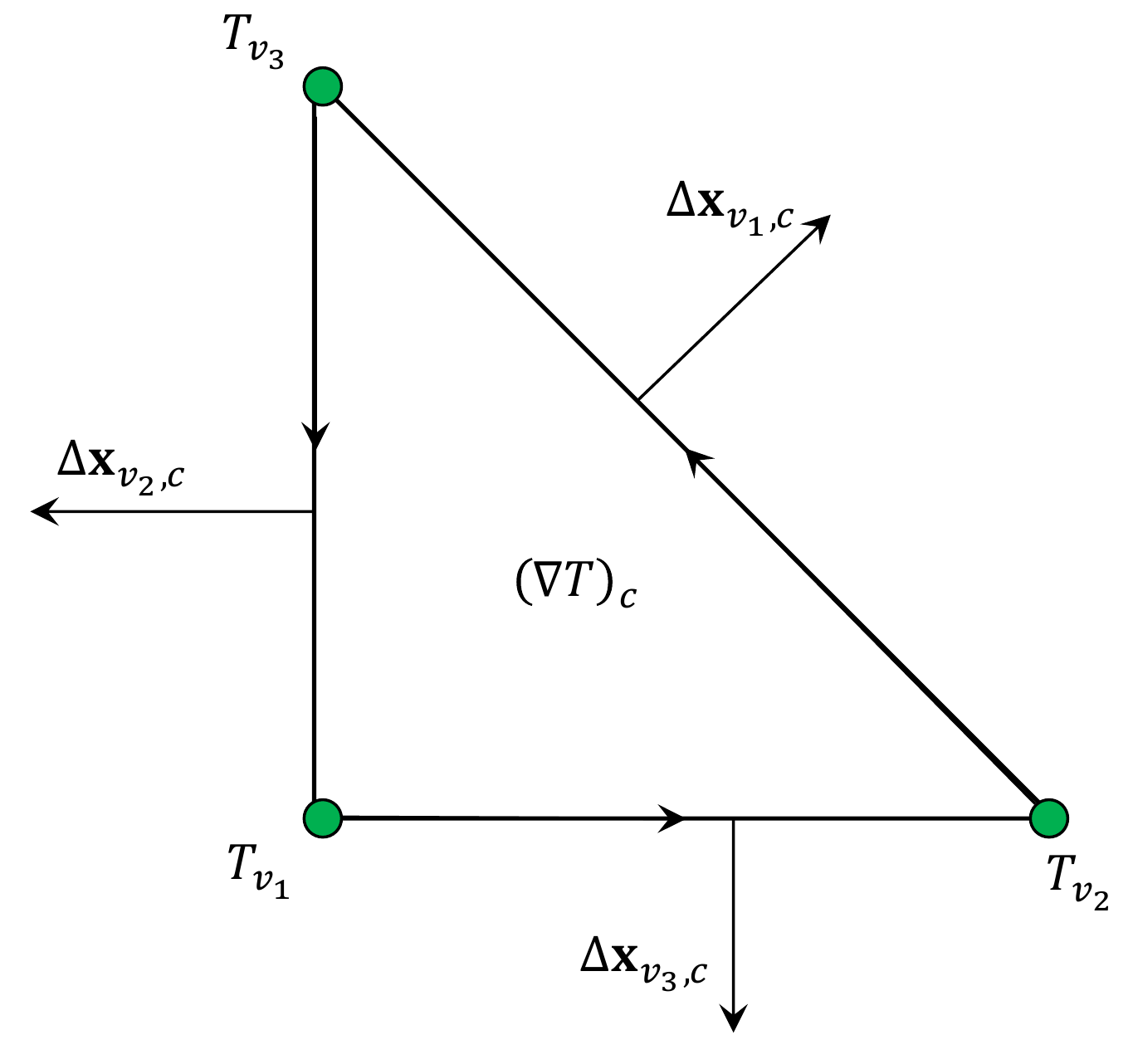}
        \captionsetup{width=0.9\linewidth}
        \caption{Definition of scalar gradient for finite volume methods on Arakawa-B grids. Scalar quantities such as temperature $T$ are defined at vertices and so scalar gradients are defined at cell centres by means of an integral over the cell and an application of Stokes' Theorem.}
        \label{fig:grad_def}
    \end{subfigure}
    \caption{Definitions of divergence and gradient for finite-volume methods with Arakawa-B placement.}
    \label{fig:meshdefs}
\end{figure}

\begin{figure}[H]
    \centering
        \includegraphics[width=.5\textwidth]{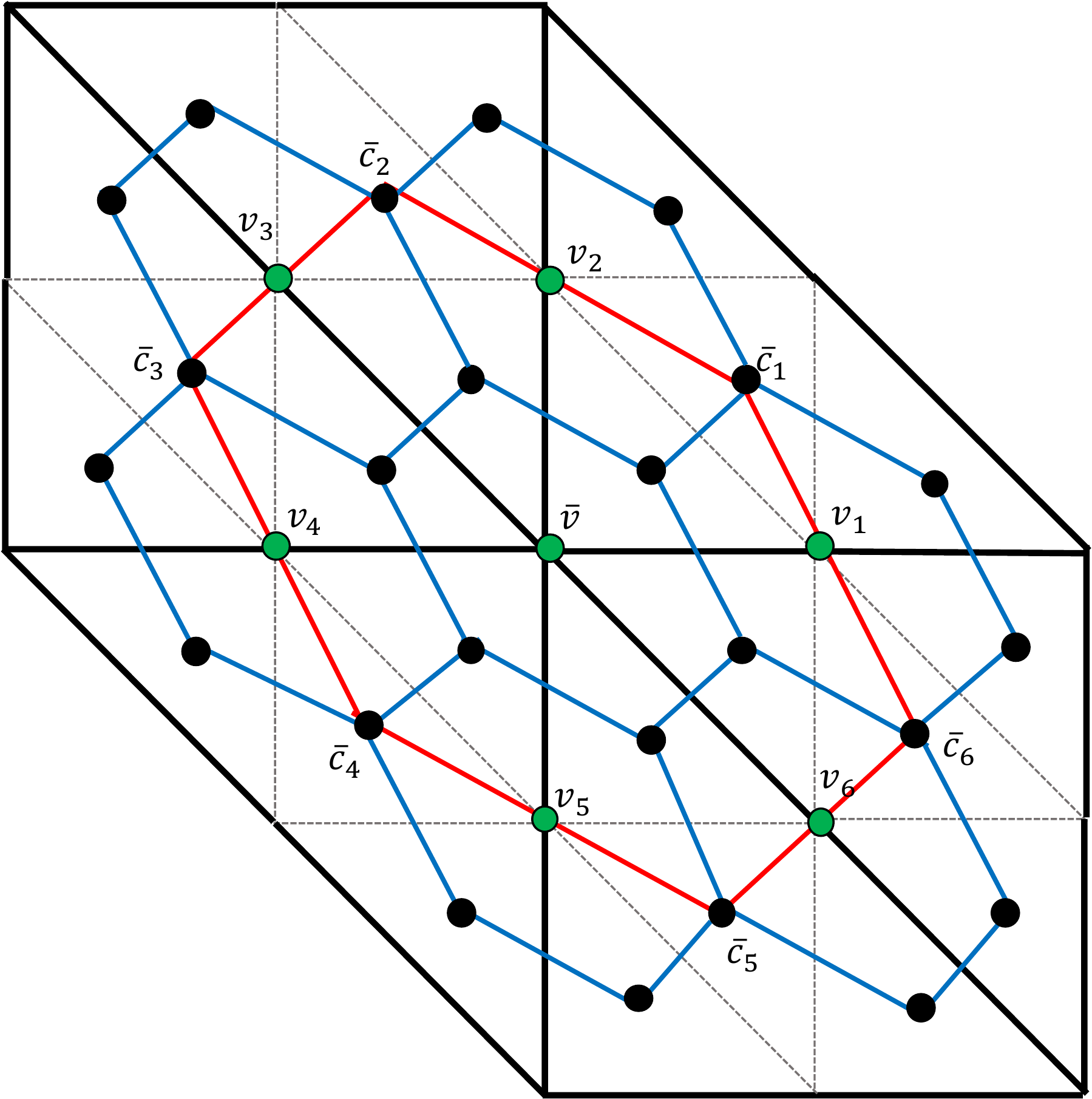}
         \captionsetup{width=0.9\linewidth}
        \caption{Averaging to preserve divergence in the case $N=2$. We calculate the averaged divergence $\lrc{\nabla\cdot \B{u}}^\al$ at coarse-grid vertex $\bar{v}$ by computing the individual fine-grid divergences at each of the green vertices (via an integral over the blue vertex cell boundary), then averaging. We compute the coarse-grid divergence $\bar{\nabla}\cdot\EXP{\B{u}}^\be$ by averaging velocities onto coarse-grid cells then integrating around the red line to get the coarse-grid divergence.}
        \label{fig:mesh2}
\end{figure}

The discretisation of the mesh used in the present study is based on that of FESOM2 \citep{danilov_sidorenko_wang_jung_2016}. Therefore we consider a triangular mesh with a placement of variables according to an Arakawa-B \citep{ARAKAWA_1977} grid. This means that scalar quantities (temperature, sea surface height, vertical velocity, etc) are defined at vertices, while vector quantities (horizontal velocity, pressure gradient, etc) are defined at cell centres, as shown in \cref{fig:meshdefs}. Unless stated otherwise, we shall assume for the purposes of this work that the grid has a regular discretisation in the horizontal directions so that vertices are spaced equally. On the fine grid, which we shall call $\mathcal{G}$, the spacing shall be taken to be $\Delta x$, while on the coarse grid $\bar{\mathcal{G}}$ we suppose the separation is $N\Delta x$ for some integer $N\geq 2$. The same vertical discretisation is used for both grids. The fine grid consists of a set of vertices $\mathcal{V}$ connected by edges, and a set of cells $\mathcal{C}$. The corresponding sets on the coarse grid will be denoted with an over-bar, so that the sets of vertices and cells in the coarse grid are $\bar{\mathcal{V}}$  and $\bar{\mathcal{C}}$ respectively. It is assumed that every vertex in the coarse grid is also a vertex in the fine grid, and that every cell in the coarse grid is the union of $N^2$ cells in the fine grid. The $N=2$ case is shown in \cref{fig:mesh2}, in which each coarse-grid cell contains $N^2 = 4$ fine-grid cells.\\
\par
A \emph{scalar field} $\phi$ on $\mathcal{G}$ is a map $\phi: \mathcal{V}\arr\mathbb{R}$. We shall denote the field $\phi$ evaluated at vertex $v$ as $\phi_v$. Let $\mathcal{F}$ denote the set of all scalar fields on $\mathcal{G}$ and let $\bar{\mathcal{F}}$ be the set of scalar fields on $\mathcal{G}$. We define a \emph{coarse-graining} to be a map $\lrc{\cdot}^{\al} : \mathcal{F}\arr\bar{\mathcal{F}}$ which sends $\phi\mapsto\lrc{\phi}^\al$ such that $\lrc{\phi}^\al$ is a scalar field on the coarse grid which is specified by means of an average over cells on the fine grid:
\eq{
    \lrc{\phi}^\al_{\bar{v}}  :=  \sum_{v\in\mathcal{V}} \al(\bar{v},v)\phi_{v} \label{eq:al_def}
}
where $\bar{v}$ is a coarse-grid vertex and $v\in\mathcal{V}$ are fine-grid vertices. $\al(\ol{v},v)$ are averaging weights and for this to be a `good' coarse-graining, $\al$ must satisfy certain properties. In particular, $\lrc{\phi}^\al$ should be a good approximation to $\phi$. If we suppose that $\phi$ is a discrete approximation of some smooth scalar field then we can do a Taylor expansion of $\phi_v$ around vertex $\bar{v}$ to find:
\algn{
     \lrc{\phi}^\al_{\bar{v}} = \sum_{v\in\mathcal{V}} \al(\bar{v},v)\lrsq{ \phi_{\bar{v}} + \lrb{\B{x}_v - \B{x}_{\bar{v}}}\cdot\lrb{\nabla \phi}_{\bar{v}} + \mathcal{O}\lrb{ \abs{\B{x}_v - \B{x}_{\bar{v}}}^2 } } \label{eq:al_approx}
}
where $\B{x}_v$, $\B{x}_{\bar{v}}$ are the coordinates of fine-grid vertex $v$ and coarse-grid vertex $\bar{v}$ respectively. So, in order to have an approximation up to second order, $\al$ should satisfy the following conditions:
\begin{subequations}
\algn{
\sum_{v\in\mathcal{V}} \al(\bar{v},v) &= 1 \label{eq:al_conditions} \\
    \sum_{v\in\mathcal{V}} \al(\bar{v},v)\B{x}_v & = \B{x}_{\bar{v}} \label{eq:al_conditionsi}
}
 The averaging weights should also decay as $v$ gets further from $\bar{v}$. Thus we define $\mathcal{V}(\bar{v}) := \lrc{v\in\mathcal{V}: \al(\bar{v},v)\neq 0}$, which will be some set of vertices neighbouring $\bar{v}$ in the fine grid. Moreover, in order for $\lrc{\cdot}^\al$ to be an average the averaging weights should all be positive:
\algn{
   \al(\bar{v},v) &\geq 0 \label{eq:al_conditions_posdef}
   }
Another property of coarse-graining that may be advantageous is to preserve integrals over the domain. This is the method used in \citep{mana2014toward} and imposes the following condition on the averaging weights:
 \algn{
    \sum_{\bar{v}\in\bar{\mathcal{V}}} \abs{\bar{A}_{\bar{v}}}\al(\bar{v},v) & = \abs{A_v} \label{eq:al_area}
 }
 for all fine-grid vertices $v$, where $A_{v}$ the vertex cell as defined in \citep{danilov_sidorenko_wang_jung_2016}. i.e. it is the region enclosed by the boundary formed by connecting cell centres to the midpoints of the adjacent edges, see \cref{fig:div_def}. The magnitude of this area is given by $\abs{A_v} = \frac{1}{3}\sum_{c\in\mathcal{C}(v)}\abs{A_c}$, where $\mathcal{C}(v)$ is the set of cells surrounding vertex $v$ and $\abs{A_c}$ is the area of cell $c$. We may observe that \cref{eq:al_area}, when combined with \cref{eq:al_conditions} implies that  $\sum_{\bar{v}\in\bar{\mathcal{V}}} \abs{\bar{A}_{\bar{v}}} = \sum_{v\in\mathcal{V}}\abs{A_v}$, i.e. the total volumes of the two domains are equal. However, this is not always the case; for example, if the boundaries of the fine and coarse grids do not exactly coincide, or if the mesh is on a curved space then the sums of all vertex areas may not be equal in the two grids. We may therefore consider replacing \cref{eq:al_conditions} with the following:
 \algn{
    \sum_{v\in\mathcal{V}}\al(\bar{v},v) &= \frac{1}{\abs{\bar{A}_{\bar{v}}}}\sum_{v\in\mathcal{V}} \kappa(\bar{v},v) \abs{A_v}  \tag{\ref*{eq:al_conditions}$'$} \label{eq:al_sum1_new}
 }
 \end{subequations}
where $\kappa$ satisfies $\sum_{\bar{v}\in\bar{\mathcal{V}}}\kappa(\bar{v},v) = 1$. \Cref{eq:al_sum1_new} is a necessary, but not sufficient condition for $\lrc{\cdot}^{\al}$ to be an integral-preserving coarse-graining. Further, it reduces to \cref{eq:al_conditions} if $\kappa$ also obeys $\sum_{v\in\mathcal{V}}\kappa(\bar{v},v)\abs{A_v}  = \abs{\bar{A}_{\bar{v}}} $.
Thus the modified condition can be thought of as an alternative to \cref{eq:al_conditions} that allows preservation of the total integral. However, using \cref{eq:al_sum1_new} will not in general keep the coefficient of $\phi_{\bar{v}}$ equal to $1$ in \cref{eq:al_approx}. The choice of which condition to use will therefore depend on the required application. \\
 \par
Similarly, we may define a \emph{vector field} $\B{v}$ on $\mathcal{G}$ to be a map $\B{v}:\mathcal{C} \arr \mathbb{R}^2$. The vector field evaluated at cell $c$ shall be denoted $\B{v}_c$ and we write $ \mathbb{V}$ to be the set of all vector fields on $\mathcal{G}$. The corresponding sets for $\bar{\mathcal{G}}$ shall be denoted with an over-bar. Thus, we define a coarse-graining on vector quantities to be a map $\EXP{\cdot}^\beta: \mathbb{V}\arr\bar{\mathbb{V}}$  such that $\EXP{\B{v}}^\be$ is a vector field on the coarse grid given by:
\eq{
 \EXP{\B{v}}_{\bar{c}}^\be  =  \sum_{c\in\mathcal{C}} \be(\bar{c},c)\B{v}_{c} \label{eq:be_def}
}
In this case the averaging weights $\beta(\bar{c},c)$ are $2\times 2$ matrices and the conditions analogous to \cref{eq:al_conditions,eq:al_conditionsi,eq:al_conditions_posdef} are:
\begin{subequations}
\algn{
  \sum_{c\in\mathcal{C}}  \be(\bar{c},c) &= \mathbb{I} \label{eq:be_conditions}\\
            \sum_{c\in\mathcal{C}} \be_{ij}(\bar{c},c) \B{x}_{c} & = \B{x}_{\bar{c}}\delta_{ij} \label{eq:be_conditionsi}\\
        \be(\bar{c},c) &\geq 0  \label{eq:be_conditions_posdef} 
        }
\Cref{eq:be_conditions_posdef} is to be interpreted as saying that $\be(\bar{c},c)$ is symmetric and positive semi-definite. $\B{x}_c$ is the coordinate of the centre of the fine grid cell $c$, $\B{x}_{\bar{c}}$ is the coordinate of the centre of the coarse grid cell $\bar{c}$ and $\delta_{ij}$ is the Kronecker delta. We denote the set of cells on which $\beta(\bar{c},\cdot)$ is non-zero as $\mathcal{C}(\bar{c}):=\lrc{c\in\mathcal{C}: \beta(\bar{c},c)\neq 0}$. In order for $\EXP{\cdot}^\be$ to be an integral-preserving coarse-graining the weights would need to satisfy:
\algn{
     \sum_{\bar{c}\in\bar{\mathcal{C}}} \abs{\bar{A}_{\bar{c}}}\be(\bar{c},c)  &= \abs{A_c} \mathbb{I} \label{eq:be_area}
}
In order for this to hold everywhere we also need to impose the following condition, analogous to \cref{eq:al_sum1_new}:
\algn{
    \sum_{c\in\mathcal{C}}\be(\bar{c},c) &= \frac{1}{\abs{\bar{A}_{\bar{c}}}}\sum_{c\in\mathcal{C}} \lambda(\bar{c},c)\abs{A_c} \tag{\ref*{eq:be_conditions}$'$} \label{eq:be_sum1_new}
}
where $\lambda(\bar{c},c)$ are $2\times2$ matrices which satisfy $\sum_{\bar{c}\in\bar{\mathcal{C}}}\lambda(\bar{c},c) = \mathbb{I}$.  If, additionally, $\lambda$ obeys $\sum_{c\in\mathcal{C}}\lambda(\bar{c},c)\abs{A_c}  = \abs{\bar{A}_{\bar{c}}}\mathbb{I} $ then  \cref{eq:be_sum1_new} reduces to \cref{eq:be_conditions}. Again, which of these we use will depend on the desired application.
\end{subequations}
\\
\par
In order to define the divergence, we follow the method used for FESOM2, as described in \cite{danilov_sidorenko_wang_jung_2016}. In particular, since this is a finite-volume \citep{barth2018finite,DANILOV201214} model we have a constant velocity $\B{v}_{c}$ on each cell, and it is assumed that the divergence at vertex $v$ multiplied by the area of the corresponding vertex cell $A_{v}$ is approximately equal to the integral of the divergence over this cell. The divergence theorem applied to the integral over this region, gives the divergence as an integral over the boundary of the vertex cell, as shown in \cref{fig:div_def}. Therefore the divergence is calculated as:
\algn{\label{eq:div_def}
    \begin{split}
    \lrsq{\nabla\cdot\B{v}}_{v} & \approx \frac{1}{\abs{A_{v}}} \int_{A_{v}} \nabla\cdot\B{v} \, dA \\
    & = \frac{1}{\abs{A_{v}}} \oint_{\pr A_{v}} \B{v}\cdot\B{n}  \, dl \\
    & = \frac{1}{2\abs{A_{v}}}\sum_{c\in\mathcal{C}(v)}\B{v}_{c}\cdot\Delta \B{x}_{v,c} 
    \end{split}
}
where $\mathcal{C}(v)$ is the set containing the six cells surrounding vertex $v$ and $\Delta \B{x}_{v,c}$ is a vector which is outward-pointing, orthogonal to the edge opposite vertex $v$ in cell $c$, and of magnitude equal to the length of that edge. The coarse-grid divergence $\bar{\nabla}\cdot $ is defined with a formula analogous to that in \cref{eq:div_def}. We remark here that in FESOM2 the areas $\abs{A_c}$ and the vectors $\Delta \B{x}_{v,c}$ vary with latitude in order to account for the curvature of the earth. We comment on the case with curvature in \cref{sec:curvature}; however, for the remainder of this section we shall consider the case in which they are constant over the grid, unless stated otherwise. \\
\par
With these definitions in mind we say that a coarse graining $\EXP{\cdot}^\be$ on vector quantities is \emph{divergence-preserving} at $\bar{v}\in\bar{\mathcal{V}}$ if there exists a coarse-graining $\lrc{\cdot}^\al$ on scalar quantities such that for any vector field $\B{v}$ defined at cell centres on the fine grid:
\algn{
    \lrsq{\bar{\nabla}\cdot \EXP{\B{v}}^\be}_{\bar{v}} = \lrc{\nabla \cdot \B{v}}^\al_{\bar{v}} \label{eq:div_condition}
}
In \cref{sec:DP_derivation} we shall find averaging weights $\al$,$\be$ such that the corresponding coarse-grainings obey \cref{eq:div_condition} for all vector fields $\B{v}$.\\
\par

Here we remark that the definition of the vertical component of the curl in FESOM2, $\PD{u_2}{x} - \PD{u_1}{y}$, is almost identical to the definition of divergence \cref{eq:div_def}, but instead of $\Delta \B{x}_{v,c}$ there is another vector of the same magnitude, pointing along the direction of integration rather than normal to it. The explicit value of these vectors will not be important, only the fact that they point in one of three directions (up to a minus sign) and sum to zero. Therefore, any coarse-graining that preserves divergence will also preserve the $z$-component of the curl. \\
\par
We also use the definition of gradient from FESOM2; this is calculated at cell $c$ by integrating over the cell, and then applying Stokes' Theorem to get an integral along edges, as shown in \cref{fig:grad_def}. The value of the scalar field at each edge is obtained by averaging the field from the vertices attached to the edge:
\begin{align}
\begin{split} \label{eq:gradient_definition}
    \lrsq{\nabla \phi}_{c} &\approx \frac{1}{\abs{A_{c}}}\int_{A_{c}} \nabla \phi \, dA  \\
    & = \frac{1}{\abs{A_{c}}}\oint_{\pr A_{c}}  \phi \B{n} \, dl \\
    & = \frac{1}{2\abs{A_{c}}}\sum_{v\in \mathcal{V}(c)} \phi_{v} \Delta \B{x}_{v,c}
\end{split}
\end{align}
where $A_{c}$ is the area of cell $c$, $\B{n}$ is the outward-pointing normal, and $\mathcal{V}(c)$ is the set of vertices attached to cell $c$. The gradient $\bar{\nabla}$ on the coarse grid is defined similarly. As with the divergence, the gradient in FESOM2 also accounts for the curvature, though the present work only considers the flat space case. \\
\par
We say that a coarse-graining $\lrc{\cdot}^{\al'}$ on scalar quantities is \emph{gradient-preserving} at $\bar{c}\in\bar{\mathcal{C}}$ if there exists a coarse-graining $\EXP{\cdot}^{\be'}$ on vector quantities such that:
\algn{
    \lrsq{\bar{\nabla}\lrc{\phi}^{\al'}}_{\bar{c}} = \EXP{\nabla \phi}^{\be'}_{\bar{c}} \label{eq:grad_condition}
}
 In \cref{sec:GP_derivation} we shall find averaging weights $\al', \be'$ such that the corresponding coarse-grainings obey \cref{eq:grad_condition} for all scalar fields $\phi$ defined on the fine grid.  We then verify numerically that our proposed coarse-graining is successful in reducing the error.

\section{Divergence-Preserving Coarse-Graining}
\label{sec:DP}
 \subsection{Derivation of Divergence-Preserving Coarse-Graining}\label{sec:DP_derivation}
With the definitions given in \cref{sec:defs}, we may derive a condition on the averaging weights for $\EXP{\cdot}^\be$ to be a divergence-preserving coarse-graining. Combining \cref{eq:div_def} with \cref{eq:div_condition} gives us:
\algn{ 
\label{eq:filt_div}
   \sum_{v\in\mathcal{V}} \al(\bar{v},v)\frac{1}{2\abs{A_{v}}}\sum_{c\in\mathcal{C}(v)}\B{v}_{c}\cdot\Delta \B{x}_{v,c}  & =     \frac{1}{2\abs{\bar{A}_{\bar{v}}}}\sum_{\bar{c}\in\bar{\mathcal{C}}(\bar{v})} \sum_{c\in\mathcal{C}}\be(\bar{c},c)\B{v}_{c}\cdot\Delta \bar{\B{x}}_{\bar{v},\bar{c}} 
}
and \cref{eq:filt_div} must hold for all vector fields $\B{v}$. We therefore re-arrange the sums in order to equate coefficients of $\B{v}_{c}$. On the left-hand side of \cref{eq:filt_div} we sum firstly over all fine-grid vertices $v\in \mathcal{V}$; then for each such vertex $v$ we sum over all cells attached to that vertex. This is equivalent to summing over all fine-grid cells $c$ then over all vertices in $\mathcal{V}$ that border the cell $c$. On the right-hand side the two sums are independent and therefore we can simply switch them. Thus:
\algn{ 
\begin{split}
\label{eq:filt_div_re-arrange}
  \sum_{c\in \mathcal{C}}\sum_{\lrc{v\in \mathcal{V}:  c\in\mathcal{C}(v)}} \frac{\abs{\bar{A}_{\bar{v}}}}{\abs{A_{v}}} \alpha(\bar{v},v)\B{v}_{c}\cdot\Delta \B{x}_{v,c} 
  = \sum_{c\in \mathcal{C} }\sum_{\bar{c}\in \bar{\mathcal{C}}(\bar{v})} \be(\bar{c},c)\B{v}_{c}\cdot \Delta \bar{\B{x}}_{\bar{v},\bar{c}}
  \end{split}
}
 Equating coefficients of $\B{v}_{c}$ in \cref{eq:filt_div_re-arrange}, we find that $\EXP{\cdot}^{\be}$ is divergence-preserving at vertex $\bar{v} \in\bar{\mathcal{V}}$ if there exists a coarse-graining $\lrc{\cdot}^{\al}$ such that for all $c\in\mathcal{C}$: 
\algn{
    \sum_{\lrc{v\in \mathcal{V}:  c\in\mathcal{C}(v)}} \frac{\abs{\bar{A}_{\bar{v}}}}{\abs{A_{v}}} \alpha(\bar{v},v)\Delta \B{x}_{v,c} & = \sum_{\bar{c}\in \bar{\mathcal{C}}(\bar{v})} \be(\bar{c},c)^{T}\Delta \bar{\B{x}}_{\bar{v},\bar{c}} \label{eq:DP_general}}
\Cref{eq:DP_general} is quite general and even holds in the case of irregular meshes, the cells of which are of any convex shape (in that case $\Delta \B{x}_{v,c}$ is orthogonal to the vector connecting the two vertices of cell $c$ that share an edge with $v$).\\
\par
Here, however, we only consider regular triangular meshes. Let us suppose that the vertex $\bar{v}$ is situated at some point in the domain which is far from the boundary so that we may assume a translational symmetry in the grid. Therefore, $\beta(\bar{c},c)$ should depend only on the relative position of the centres of cells $\bar{c}$ and $c$, not on the absolute position within the grid. Similarly, $\al\lrb{\bar{v},v}$ should not depend on the position of $\bar{v}$, only the relative position of the vertices $\bar{v}$ and $v$. \\ \par
It is convenient to define a metric $d:\mathcal{G}\times\mathcal{G}\arr\mathbb{N}\cup \lrc{0}$ on vertices such that $d(v,v')$ is a integer equal to the minimum number of fine-grid edges we must traverse in order to arrive at vertex $v'$ starting from vertex $v$. Because of the translational symmetry we suppose $\al(\bar{v},v)$ depends only on $d(\bar{v},v)$ and therefore we write $\al(\bar{v},v) =: \al_{d(\bar{v},v)}$. Further, we assume that $\al_{n} = 0$ for $n\geq N$; that is, we set $\mathcal{V}(\bar{v}) = \lrc{v\in\mathcal{V} : d(\bar{v},v)<N}$. Then there are $N$ variables $\al_{0},\al_1,...,\al_{N-1}$ to determine. We also specify $\mathcal{C}(\bar{c})$ by setting $\be(\bar{c},c)=0$ unless $c$ is contained in $\bar{c}$. With this assumption the sum on the right-hand side of \cref{eq:DP_general} has only one term.\\
\par
 Since the mesh is regular and triangular there are  (up to a minus sign) only three possible values for $\Delta \B{x}_{v,c}$. We denote these vectors $\Delta \B{x}_{i}$ and note that $\sum_{i=1}^3 \Delta \B{x}_i = 0$. On the coarse-grid the corresponding vectors are simply rescaled by the ratio of resolutions: $\Delta \bar{\B{x}}_i = N\Delta \B{x}_i$. \\
 \par
\begin{figure}[H]
    \centering
        \includegraphics[width=.5\textwidth]{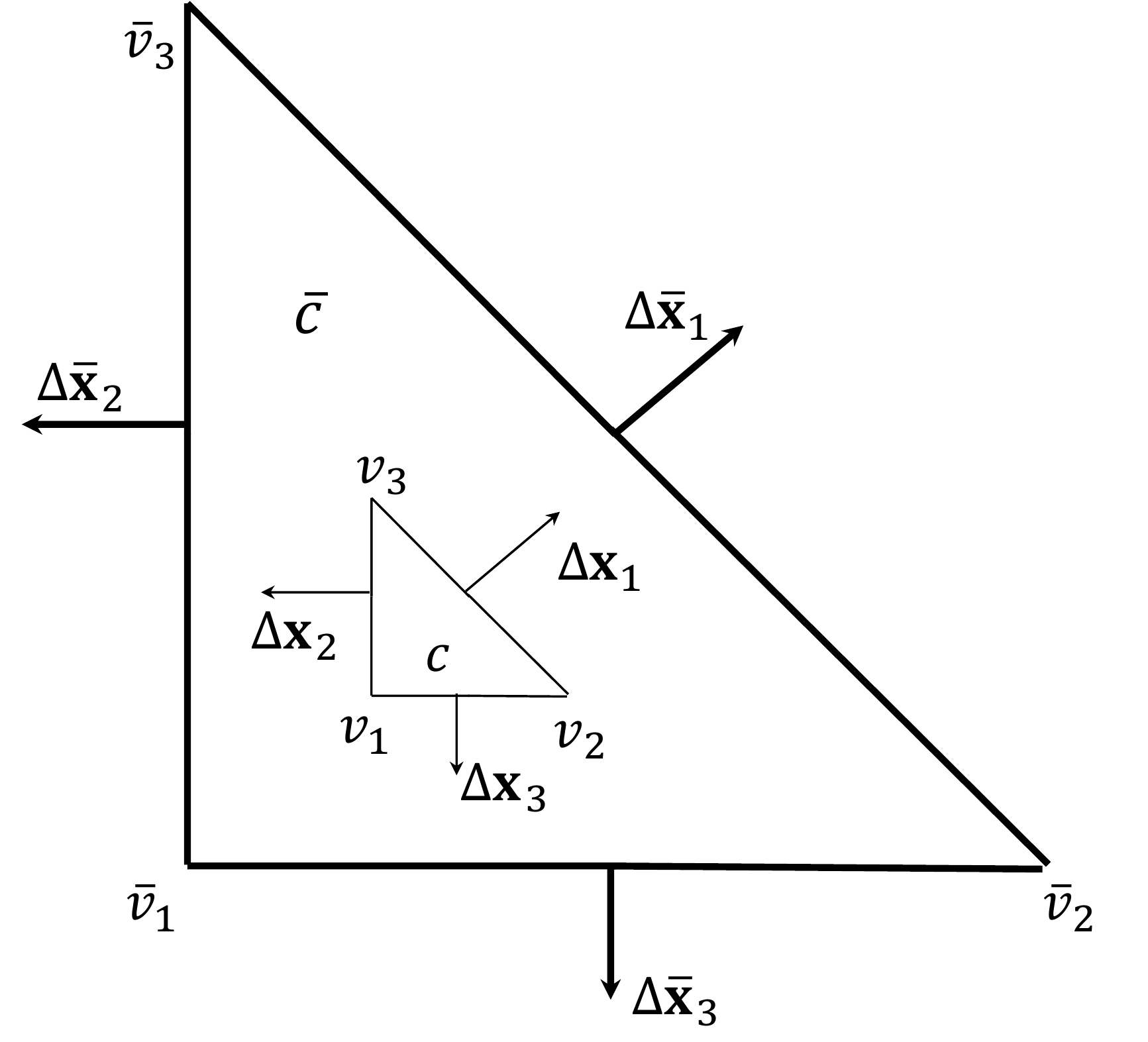}
         \captionsetup{width=0.6\linewidth}
        \caption{Fine grid cell contained within coarse grid cell for general $N$. \Cref{eq:DP_general} must hold for every fine-grid cell $c$ and every coarse-grid vertex $\bar{v}$. If we take $\be\lrb{\bar{c},c}$ to be non-zero only if $c$ is contained in $\bar{c}$ then we have three equations for each $\be(\bar{c},c)$, corresponding to each of the coarse-grid vertices of $\bar{c}$.}
        \label{fig:DP_N}
\end{figure}

Suppose $c$ is an upward-pointing fine-grid cell contained in an upward-pointing coarse-grid cell $\bar{c}$, as shown in \cref{fig:DP_N}. And let us consider the coarse-grid vertices $\bar{v}_{i}$ in turn. Then \cref{eq:DP_general} gives, for $i=1,2,3$:
\algn{
\begin{split} \label{eq:beta_eq_gen}
    \be(\bar{c}, c)^T \Delta \B{x}_{i}  = &N\al\lrb{\bar{v}_{i},v_{1}}\Delta \B{x}_{1}  +  N\al\lrb{\bar{v}_{i},v_{2}}\Delta \B{x}_{2}  + N\al\lrb{\bar{v}_{i},v_{3}}\Delta \B{x}_{3}
\end{split}
}
where we have used the fact that on a regular grid with $\bar{v}, v$ away from the boundary the ratio of areas is $\abs{\bar{A}_{\bar{v}}}/\abs{A_{v}} = N^{2}$. Further, we notice that since $c$ is an upward-pointing triangle, we have $d(\bar{v}_{i},v_{i\pm1}) = d(\bar{v}_{i},v_{i}) + 1$. We may therefore use the fact that $\sum_{i=1}^{3}\Delta \B{x}_i = 0 $ to write:
\algn{
     \be(\bar{c}, c)^T \Delta \B{x}_{i} & = N\lrb{\al_{d(\bar{v}_{i},v_{i})} - \al_{d(\bar{v}_{i},v_{i})+1}}\Delta \B{x}_{i}  \label{eq:be_i}
}
We obtain an identical result when $c$ is a downward-pointing triangle. Summing \cref{eq:be_i} over $i=1,2,3$ allows us to deduce that:
\algn{
    \al_{d(\bar{v}_{i\pm1},v_{i\pm1})} - \al_{d(\bar{v}_{i\pm1},v_{i\pm1})+1}  = \al_{d(\bar{v}_{i},v_{i})} - \al_{d(\bar{v}_{i},v_{i})+1} \label{eq:all_c}
}
\Cref{eq:all_c} holds for $v_{i},v_{i\pm1}$ being the vertices of any cell $c$ contained in the cell $\bar{c}$ with vertices $\bar{v}_{i}, \bar{v}_{i\pm 1}$. A direct consequence of \cref{eq:all_c} when combined with \cref{eq:be_i} is that $\be(\bar{c},c)$ must all be equal and proportional to the identity. But $\mathcal{C}(\bar{c})$ contains $N^2$ elements therefore by \cref{eq:be_conditions} we find:
\algn{
    \be(\bar{c},c) = \frac{\chi_{\mathcal{C}(\bar{c})}(c)}{N^2}\mathbb{I}  \label{eq:solution_be_DP}
}
where $\chi_{\mathcal{C}(\bar{c})}(c)$ is the indicator function, equal to $1$ if $c\in\mathcal{C}(\bar{c})$ and $0$ otherwise. It is readily verified that the solution given by \cref{eq:solution_be_DP} satisfies \cref{eq:be_conditionsi}. Furthermore, we deduce from \cref{eq:be_i,eq:solution_be_DP} that for all $n\leq N$ we have:
\algn{
    \al_{n} - \al_{n+1} = \frac{1}{N^{3}}
}
Then we can use the fact that $\al_{N} = 0$ to solve inductively, giving the result, for $n<N$:
\algn{
    \al_{n}  &= \frac{1}{N^2}\lrb{1-\frac{n}{N}} \label{eq:al_solution_DP}
}
For $n>0$ there are $6n$ vertices at distance $n$ from a coarse-grid vertex $\bar{v}$, so we can verify \cref{eq:al_conditions} since  $\sum_{v\in\mathcal{V}(\bar{v})}\al(\bar{v},v) = \frac{1}{N^2}\lrb{1 + 6\sum_{n=1}^{N-1}n\lrb{1-\frac{n}{N}}  }= 1$.  \\
\par 
Moreover, since we are considering a regular grid, the areas obey $\abs{A_v}/\abs{\bar{A}_{\bar{v}}} = \abs{A_c}/\abs{\bar{A}_{\bar{c}}} = 1/N^2$ for cells and vertices located away from the boundaries. Therefore, \cref{eq:al_area} also holds for vertices $v$ away from the boundary, as does \cref{eq:al_sum1_new} with $\kappa(\bar{v},v) = \lrb{1 - \frac{d(\bar{v},v)}{N}}\chi_{\mathcal{V}(\bar{v})}(v)$ which satisfies $\sum_{\bar{v}\in\mathcal{V}}\kappa(\bar{v},v)=1$. $\be$ similarly obeys \cref{eq:be_area,eq:be_sum1_new} for $c$ away from the boundary, with $\lambda(\bar{c},c) = \chi_{\mathcal{C}(\bar{c})}(c)\mathbb{I}$. Therefore, we find that at points away from the boundaries, the solutions for $\al,\be$ given in \cref{eq:al_solution_DP,eq:solution_be_DP} define a divergence-preserving coarse-graining; these coarse-grainings also give a good approximation to the original fields, as they obey \cref{eq:al_conditionsi,eq:be_conditionsi}; and they preserve integrals, since \cref{eq:al_area,eq:be_area} also hold.
\\
\par
We may also extend our solution to the boundaries provided the boundaries of the fine and coarse grids coincide. If $\bar{v}$ is a boundary vertex in the coarse grid then its corresponding vertex area $\abs{\bar{A}_{\bar{v}}}$ will not be the same as the vertex area of a vertex located off the boundary. Therefore we use the following generalisation of \cref{eq:al_solution_DP}:
\algn{
    \al(\bar{v},v) = \lrb{1 - \frac{d(\bar{v},v)}{N}}\frac{\abs{A_v}}{\abs{\bar{A}_{\bar{v}}}} \chi_{\mathcal{V}(\bar{v})}(v)\label{eq:al_solution_gen}
}
 It can be shown using similar arguments to before that the choice given in \cref{eq:al_solution_gen} obeys \cref{eq:DP_general} with the same $\beta(\bar{c},c)$ as before. Moreover, \cref{eq:al_solution_gen} can be shown to satisfy \cref{eq:al_area,eq:al_sum1_new} with $\kappa(\bar{v},v) = \lrb{1 - \frac{d(\bar{v},v)}{N}}\chi_{\mathcal{V}(\bar{v})}(v)$; \cref{eq:al_conditions} is also satisfied on a regular grid. The averaging weights given by \cref{eq:al_solution_gen} do not however, satisfy \cref{eq:al_conditionsi}.  \\
\par
\begin{figure}[H]
    \centering
        \includegraphics[width=.5\textwidth]{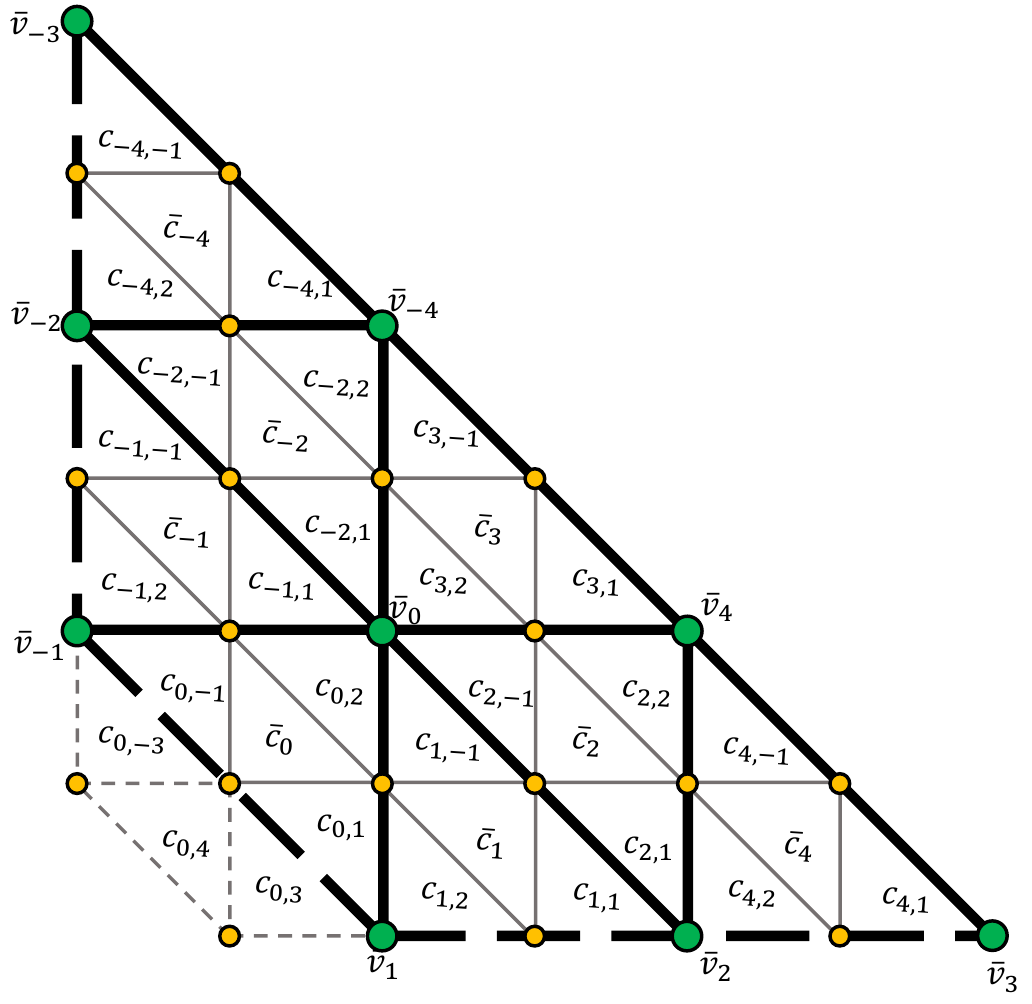}
         \captionsetup{width=0.6\linewidth}
        \caption{Corner of meshes used in FESOM2 runs for $N=2$. The thick dotted lines are the boundary of the coarse grid, while the thin dotted lines are the edges of fin-grid cells. At each corner there are three cells in the fine grid not contained in the coarse grid.}
        \label{fig:corner}
\end{figure}

A difficulty arises where the fine-grid boundary contains cells which lie outside the boundary of the coarse-grid. Such is the case for the model we use for our numerical results, in which the north-east and south-west corners of the domain are resolved differently in the fine and coarse grids; the south-western corner is shown in \cref{fig:corner} for $N=2$. The $\al,\be$ derived above will not satisfy \cref{eq:DP_general} and \cref{eq:be_conditions} at the corners because, for example, the $\al$-averaging onto vertex $\bar{v}_1$ includes vertices which lie outside the boundary of the coarse grid, while the averaging onto coarse-grid cells $\bar{c}_0, \bar{c}_{\pm1}$ does not include the cells which lie outside the coarse-grid boundary. It is unclear at present how to define $\mathcal{C}(\bar{c})$ and $\mathcal{V}(\bar{v})$ at these points such that it is possible to find $\al(\B{v},\cdot)$ and $\beta(\bar{c},\cdot)$ obeying \cref{eq:DP_general}. Further investigation is needed into the divergence-preserving problem at the boundary and this is a potential subject for future study.\\
\par
It is possible to define an integral-preserving coarse-graining on vector fields at such boundaries by, for example, choosing $\mathcal{C}(\bar{c}_0)$ to include the cells contained in $\bar{c}_0$, along with the adjacent fine-grid cells which lie outside the coarse-grid boundary. \Cref{eq:be_area} then gives $\beta(\bar{c}_0,c)=\frac{1}{N^2}\mathbb{I}$ on these cells. With this choice \cref{eq:be_sum1_new} holds with $\lambda(\bar{c}_0,c) = \mathbb{I}\chi_{\mathcal{C}(\bar{c}_0)}(c)$ and so this defines an integral-preserving coarse graining.

\subsection{Effect of Curvature}
\label{sec:curvature}
A further potential to extend the ideas explored herein would be to take into account irregular meshes, or those with curvature. As mentioned previously, the formula for computing divergences in FESOM2 is the same as that given in \cref{eq:div_def}, but now $\abs{A_c}$, $\abs{A_v}$ and $\Delta \B{x}_{v,c}$ vary with latitude. Concretely, we write $\abs{A_c} = \abs{A}\cos y_c$, where $\abs{A}$ is a constant area and $y_c$ denotes the latitude of the centre of cell $c$. We also have a distortion of the normal vectors: $\Delta \B{x}_{v,c} = D(c)\Delta \tl{\B{x}}_{v,c}$, where $D(c) = \twobytwomatrix{1 }{0}{0}{\cos y_{c}}$ and $\Delta \tl{\B{x}}_{v,c}$ are the corresponding flat- grid vectors, as used previously. It is convenient to define re-scaled $\al$ and $\be$, so we let $\tl{\al}(\bar{v},v):= \frac{\abs{\bar{A}_{\bar{v}}}}{N^2\abs{A_v}}\al(\ol{v},v)$ and $\tl{\be}(\bar{c},c) := D(\bar{c})\be(\bar{c},c)D(c)^{-1}$. By doing this we recover \cref{eq:beta_eq_gen} but with $\al\arr\tl{\al}$, $\be\arr\tl{\be}$, and $\Delta \B{x}_{v,c}\arr\Delta \tl{\B{x}}_{v,c}$. Therefore the divergence-preserving solution is given by $\tl{\al}(\bar{v},v) = \frac{1}{N^2}\lrb{1-\frac{d(\bar{v},v)}{N}}\chi_{\mathcal{V}(\bar{v})}(v)$ and $\tl{\be}(\bar{c},c) = \frac{\chi_{\mathcal{C}(\bar{c})}(c)}{N^2}\mathbb{I}$ away from the boundaries; equivalently $\al(\bar{v},v) = \frac{\abs{A_v}}{\abs{\bar{A}_{\bar{v}}}}\lrb{1-\frac{d(\bar{v},v)}{N}}\chi_{\mathcal{V}(\bar{v})}(v)$ and $\be(\bar{c},c) = \frac{\chi_{\mathcal{C}(\bar{c})}(c)}{N^2}D(\bar{c})^{-1}D(c)$. This choice of $\al$ obeys \cref{eq:al_area,eq:al_sum1_new} away from the corners, with $\kappa(\bar{v},v) = \lrb{1-\frac{d(\bar{v},v)}{N}}\chi_{\mathcal{V}(\bar{v}}(v)$. Thus, the coarse-graining defined by $\al$ is integral-preserving away from the corners. The coarse-graining defined by $\be$ obeys neither \cref{eq:be_conditions} nor \cref{eq:be_sum1_new}. We can, however, show that $\sum_{c\in\mathcal{C}(\bar{c})}\cos y_c = N^2 \cos\bar{y}_{c}\lrb{1 + \mathcal{O}(N^2 \Delta x^2)}$ and therefore $\sum_{c\in\mathcal{C}(\bar{c})}\beta(\bar{c},c) = \mathbb{I} + \twobytwomatrix{1}{0}{0}{\mathcal{O}(N^2\Delta x^2)}$, so that \cref{eq:be_conditions} is satisfied in the $x$-direction, and satisfied up to $\mathcal{O}(N^2\Delta x^2)$ in the $y$-direction. 
\\
\par
For $\EXP{\cdot}^\be$ to be integral-preserving we would need $\be(\bar{c},c) =\frac{\abs{A_c}}{\abs{\bar{A}_{\bar{c}}}}\chi_{\mathcal{C}(\bar{c})}(c)\mathbb{I}$ so that \cref{eq:be_area} is satisfied as long as for each $c\in\mathcal{C}$ there is exactly one $\bar{c}\in\mathcal{C}(\bar{c})$ with $c\in\mathcal{C}(\bar{c})$. In this case \cref{eq:be_sum1_new} also holds with $\lambda(\bar{c},c) = \chi_{\mathcal{C}(\bar{c})}(c)\mathbb{I}$. At the corners we can use the same prescription as before, averaging the cells outside the coarse boundary onto the adjacent cell inside the boundary. This solution, however, will not be divergence-preserving. Thus when coarse-graining vector fields on grids with curvature we can choose a divergence-preserving coarse-graining which approximately obeys \cref{eq:be_conditions} and which does not preserve integrals; or we can choose an integral-preserving coarse-graining which does not preserve divergences.

\subsection{ALE coordinates}

One of our motivations for defining a divergence-preserving coarse graining was in order to give the correct vertical velocity when computed on the coarse grid. However, FESOM2 may be configured to use Arbitrary Lagrangian-Eulerian (ALE) coordinates, which allows the thicknesses of the vertical layers to vary in time. Details on how ALE coordinates are implemented can be found in \citep{danilov_sidorenko_wang_jung_2016}, but here we repeat some of the relevant aspects. In particular, the vertical velocities, since they are now defined at moving vertical coordinates, are given by:
\algn{\label{eq:ALE}
    w_{k} = w_{k+1} - \pr_{t}h_{k} - \lrsq{\nabla\cdot\lrb{h\B{u}}}_k
}
where $h_k$ is the thickness of layer $k$ (defined at vertices), and $k$ increases with depth so that $h_0$ is the thickness of the surface layer, etc. The evolution of the layer thicknesses, $\pr_t h_k$ is specified as one of several options. The fixed-layer case $\pr_t h_k = 0$ we have already considered. Another popular option is to use so-called $z^*$ coordinates in which the change in the total layer thickness $\hat{h} := \sum_k h_k - H$ is distributed over all layers, so that we impose $\pr_t h_k = \frac{h_k^0}{H}\pr_{t}\hat{h}$, where $h_k^0$ is the initial thickness of layer $k$. The evolution of the total layer thickness is given by $\pr_{t}\hat{h} = -\nabla\cdot\int_{-H}^0 \B{u} dz = -\nabla\cdot\sum_k (h \B{u})_k $ and so in this case the vertical velocity is given by:
\algn{\label{eq:zstar}
    w_{k} = w_{k+1} + \frac{h_k^0}{H}\lrsq{\nabla\cdot \sum_k (h \B{u})_k } - \nabla\cdot\lrb{h\B{u}}_k
}
In the expression $h\B{u}$, $h$ must be defined at cell centres, and in FESOM2 this is calculated this via an average over surrounding vertices: $h_c := \frac{1}{3}\sum_{v\in\mathcal{V}(c)} h_v$. There are two divergences to be calculated in \cref{eq:zstar}, and we note that these may be preserved if we coarse grain the velocities according to $\bar{\B{u}}_{\bar{c}} := {\EXP{h\B{u}}}_{\bar{c}}/{\bar{h}_{\bar{c}}}$, where $\EXP{\cdot}$ is a divergence-preserving coarse-graining and $\bar{h}_{\bar{c}}$ is the layer thickness at cell $\bar{c}$ in the coarse grid. Thus by making a small modification we are able to adapt the coarse-graining method to obtain the correct vertical velocity when $z^*$ coordinates are used. \\
\par
Other versions of ALE coordinates are available, and the method for defining the coarse-grained velocity is likely to be different in those cases; indeed it may not be possible for all methods to define an appropriate coarse-graining that preserves the vertical velocity. However, we do not consider every possible choice in the present work, but merely include this as an example of how the method may be adapted.

\subsection{Numerical results} \label{sec:Num_results}

Here we take the special case $N=2$ and verify our results numerically using output from FESOM2  \citep{danilov_sidorenko_wang_jung_2016}. The definition of divergence used in the model is slightly different from that given in \cref{eq:div_def} because FESOM2 takes into account curvature and so there are additional factors of $\cos y$ (where $y$ is the latitude) in the cell and vertex cell areas, and in the normals. Therefore, we do not expect our coarse-graining to preserve the divergence exactly. For our model set-up we use a domain $[0,40^{\circ}]\times[30^{\circ},60^{\circ}]\times[0,-1600m]$ in the longitude, latitude and vertical directions respectively. For the fine-grid we use vertices spaced at intervals of $1/4^{\circ}$ in the latitude and longitude directions. For the coarse-grid we use a spacing of $1/2^{\circ}$. In both grids the vertical resolution contains 23 layers, spaced closer together towards the surface of the domain. We do not use ALE coordinates, so layer thicknesses are fixed. The flow is driven by a wind-forcing at the surface and temperature relaxation at the northern and southern boundaries. \\ \par
\begin{figure}[H]
    \centering
    \begin{subfigure}{.5\textwidth}
      \centering
        \includegraphics[width=.9\textwidth]{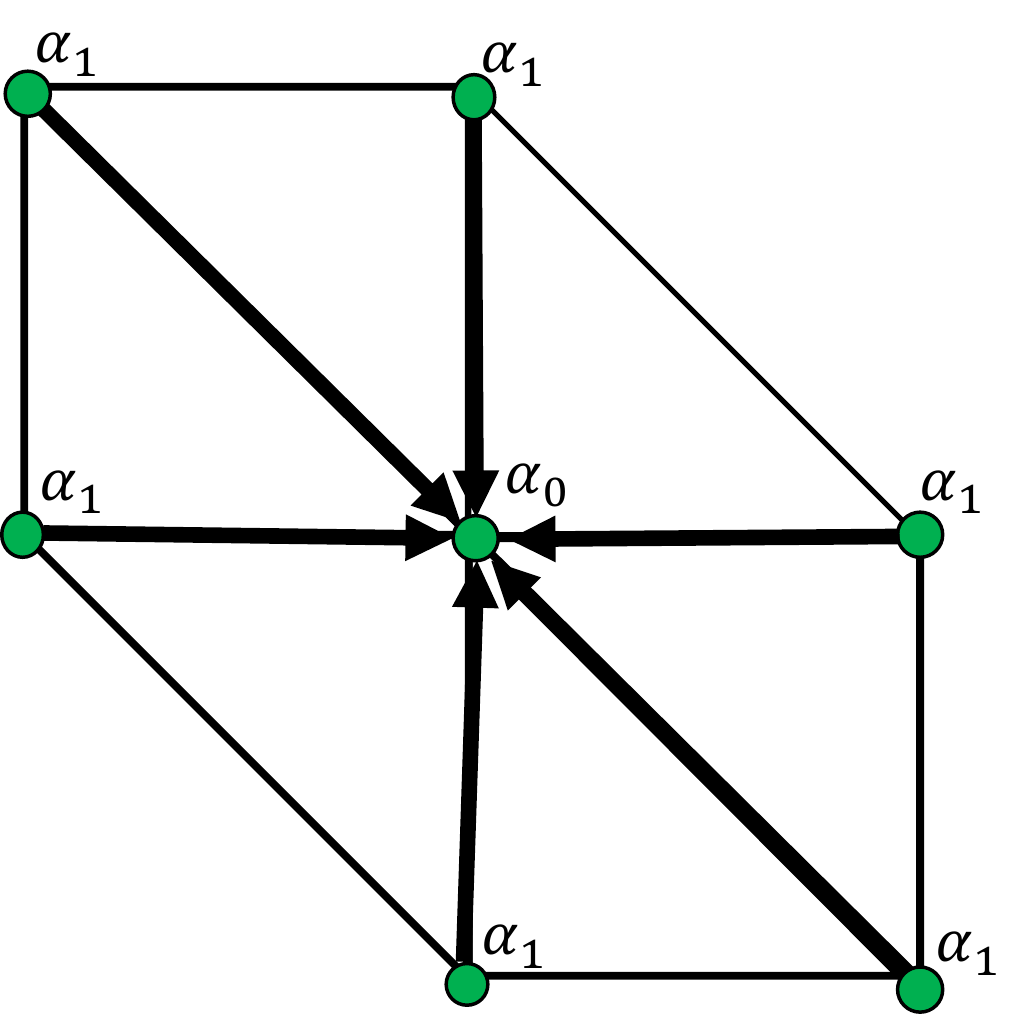}
         \captionsetup{width=0.9\linewidth}
        \caption{Seven-point average over a vertex and its immediate neighbours.}
        \label{fig:Vv}
    \end{subfigure}%
    \begin{subfigure}{.5\textwidth}
      \centering
        \includegraphics[width=.9\textwidth]{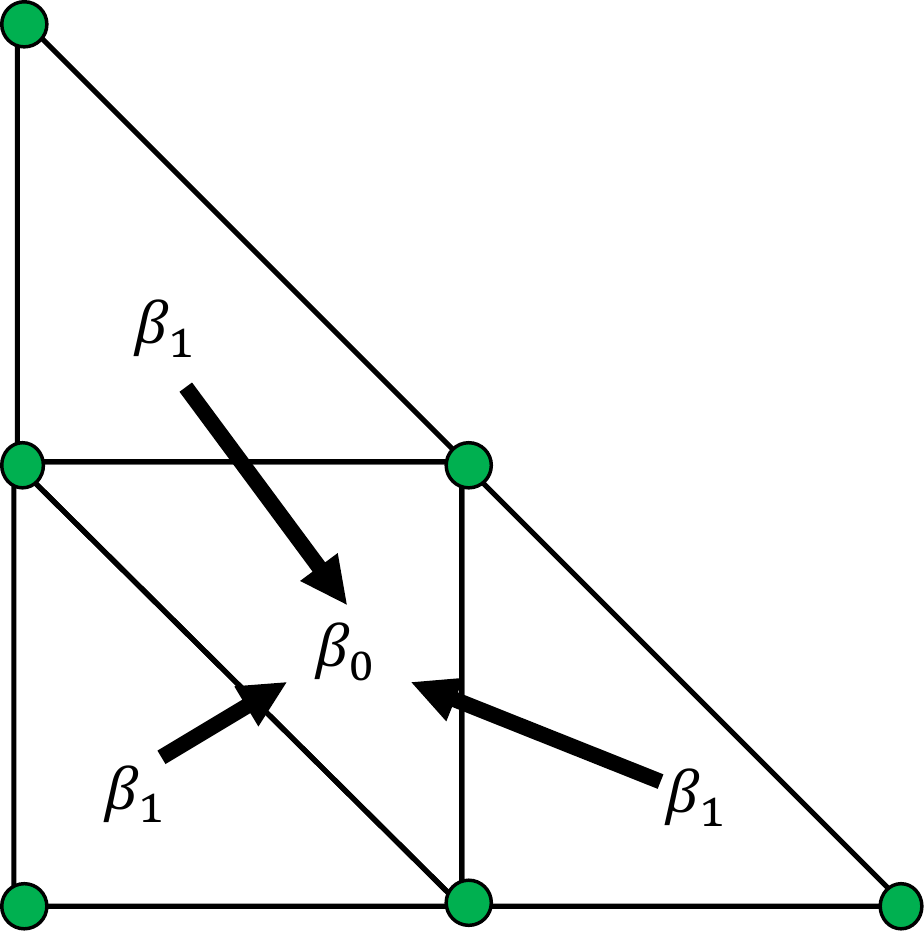}
         \captionsetup{width=0.9\linewidth}
        \caption{Four-point average on cells.}
        \label{fig:Cc}
    \end{subfigure}
    \caption{Examples of averages on vertices and cells. }
    \label{fig:mesh_averages}
\end{figure}

We run the fine-grid model for one year (after 20 years of spin-up) and save the output horizontal velocities every 5 days. We then consider a coarse-graining consisting of the four-point average shown in \cref{fig:Cc}, i.e. in order to obtain the velocity on a coarse-grid cell, we average over all fine-grid cells contained in that cell. According to the results of \cref{sec:DP_derivation}, the divergence-preserving coarse-graining will have $\be_{0}=\be_{1}=\frac{1}{4}$. Here we use a range of values: $\beta_{1} = 0,1/6,1/4,1/3$ so that $\be_{0}=1,1/2,1/4,0$ (since $\be_{0}=1-3\be_{1}$). We also consider a further coarse-graining, called the `vertex-cell averaging' (henceforth VC averaging) , which consist of averaging from cells (weighted by cell area) onto vertices, then averaging back onto cells (this is described in Appendix C of \cite{juricke_danilov_kutsenko_oliver_2019}). Furthermore, we include a divergence-preserving solution modified to take into account the curvature, i.e. $\beta(\bar{c},c) = \frac{1}{N^2}D(\bar{c})^{-1}D(c)\chi_{\mathcal{C}(\bar{c})}(c)$ with $\mathcal{C}(c)$ consisting of exactly the fine-grid cells contained in $\bar{c}$. Finally, we consider an integral-preserving coarse-graining $\beta(\bar{c},c) = \frac{\abs{A_c}}{\abs{\bar{A}_{\bar{c}}}}\chi_{\mathcal{C}(\bar{c})}(c)\mathbb{I}$ where $\mathcal{C}(\bar{c})$ consists of those fine grid cells contained in $\bar{c}$, but if $\bar{c}$ lies on a boundary such that there are fine grid cells outside the boundary of the fine grid adjacent to $\bar{c}$, then these cells are also included in $\mathcal{C}(\bar{c})$. \\
\par
\begin{figure}[H]
    \centering
    \includegraphics[width=1.0\textwidth]{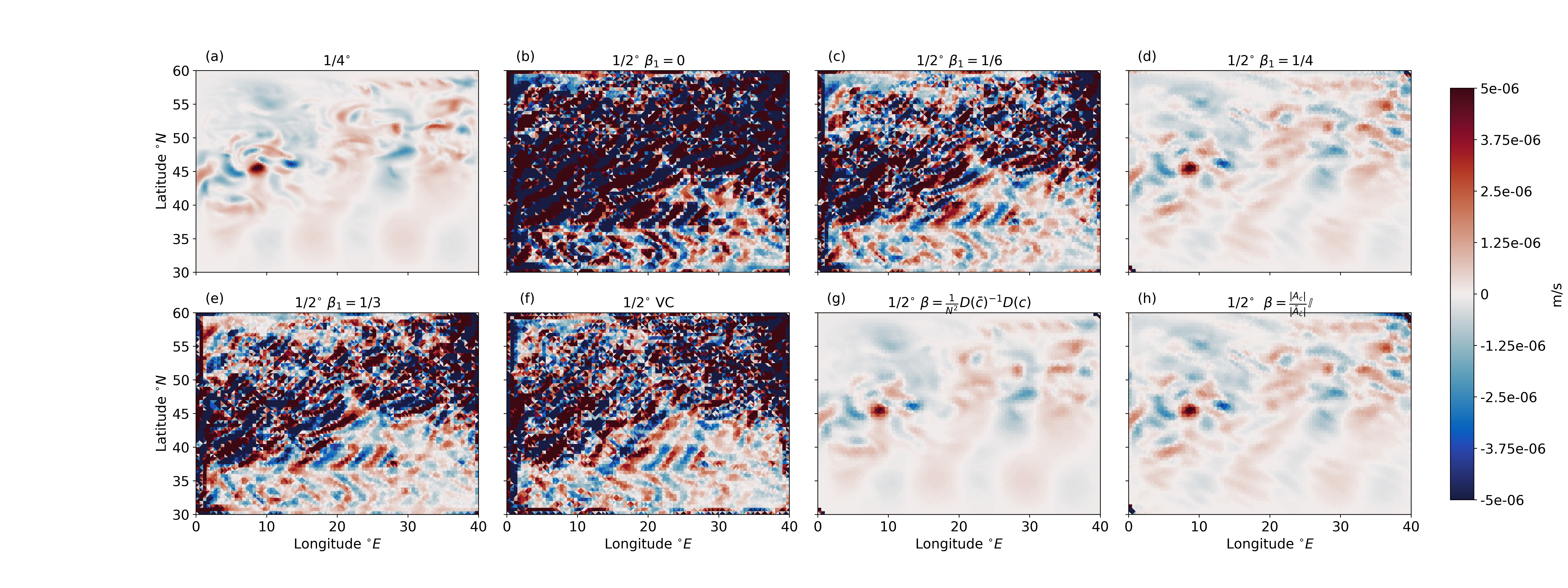}
    \caption{Snapshot of vertical velocities at the surface. Panel (a) is the reference solution on the fine grid. Panels (b)-(h) are the coarse-grid vertical velocities calculated from various applications of coarse-graining. (b)-(e) use the coarse-graining shown in \cref{fig:Cc} and (f) uses the vertex-cell averaging. (g) uses an exactly divergence-preserving coarse-graining taking the curvature of the mesh into account. (h) is an integral-preserving coarse-graining. }
    \label{fig:w_filt}
\end{figure}

After computing the coarse-grained solutions we run FESOM2 on the coarse grid and load in the coarse-grained velocities, so that the corresponding vertical velocities are calculated online. The resulting $w$ fields in the surface layer are shown in \cref{fig:w_filt}. We see that taking $\be_{1}=\be_{0}=1/4$, as we derived in \cref{sec:DP_derivation}, gives a noticeable improvement over the other values and over the VC averaging. In these latter cases we see grid effects appearing in the vertical velocities calculated on the coarse grid. However, when we average with $\be_{1}=1/4$ these effects largely disappear, leaving a field strongly resembling the reference solution. A significant discrepancy is the presence of large vertical velocities in the north-east and south-west corners of the domain; this is a result of the fact that the boundary of the fine grid differs slightly from the boundary of the coarse grid at these points. There also seem to be some artificial streaks in the domain, with some of the large-scale structures distorted. The vertical velocity field in the top layer which results from the integral-preserving choice, $\be(\bar{c},c) = \beta(\bar{c},c) = \frac{\abs{A_c}}{\abs{\bar{A}_{\bar{c}}}}\chi_{\mathcal{C}(\bar{c})}(c)\mathbb{I}$ is almost identical to the vertical velocity in the case $\be=1/4$, though the error in the north-east corner seems to have grown; the large-scale streaks are also present in this case. For the coarse-graining defined by $\beta(\bar{c},c) = \frac{1}{N^2}D(\bar{c})^{-1}D(c)\chi_{\mathcal{C}(\bar{c})}(c)$ the large-scale streaks seem to have been eliminated, indicating that this source of error was indeed caused by not taking the curvature into account. Some errors remain in the corners of the domain. \\ 
\par
\begin{figure}[H]
     \centering
     \includegraphics[width=.8\textwidth]{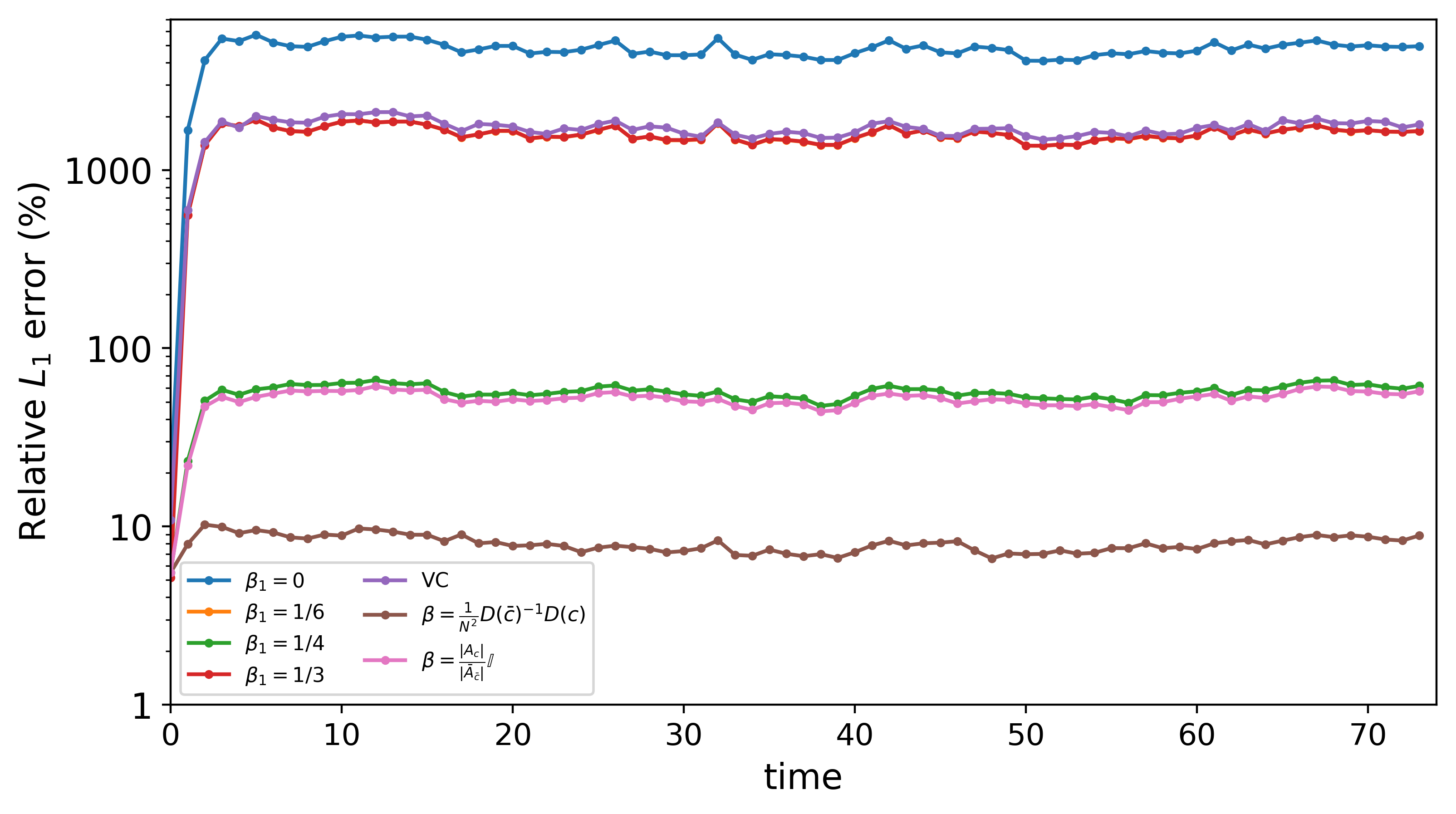}
     \caption{Relative $L_{1}$ error between $\bar{w}$ calculated from coarse-grained velocities and the fine grid $w$ projected directly onto coarse mesh. The $L_{1}$ error is given by $E[w,\bar{w}] = \frac{\int \abs{w - \bar{w}} d^{2}x}{\int \abs{w} d^{2}x}$ and in the integral we exclude points at a perpendicular distance of $1.5^{\circ}$ from the lateral boundaries. The error is plotted at the surface.}
     \label{fig:w_L1}
 \end{figure}
\begin{figure}[H]
     \centering
     \includegraphics[width=1.0\textwidth]{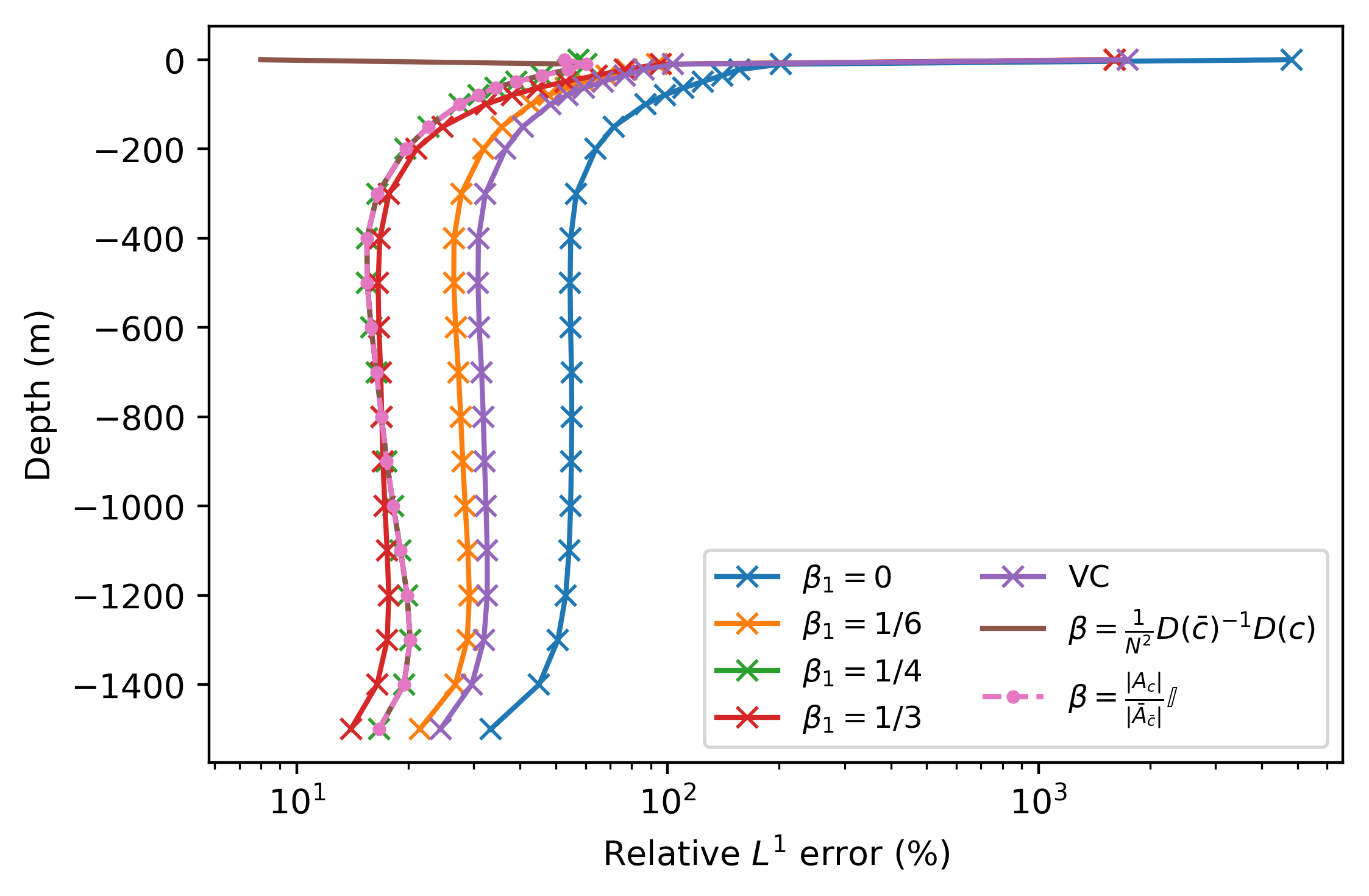}
     \caption{Time-average of relative $L^{1}$ error in vertical velocity, at different depths. Time-average excludes first five points.}
     \label{fig:w_L1_depth}
 \end{figure}

The relative $L^{1}$ errors between the coarse-grid vertical velocities in the top layer and those of the reference solution are shown in \cref{fig:w_L1}. There we see that there is a significantly smaller error in the case $\be_{1}=1/4$ than in the cases with other choices of $\be_1$ and a much smaller error than the VC averaging. However, a slightly smaller error is achieved with the integral-preserving coarse-graining, while a significant reduction in error occurs in the coarse-graining taking curvature into account. \Cref{fig:w_L1_depth} shows that, apart from in the top layer, the difference between $\be_{1}=1/4$, $\beta(\bar{c},c) = \frac{1}{N^2}D(\bar{c})^{-1}D(c)\chi_{\mathcal{C}(\bar{c})}(c)$ and $\be(\bar{c},c) = \beta(\bar{c},c) = \frac{\abs{A_c}}{\abs{\bar{A}_{\bar{c}}}}\chi_{\mathcal{C}(\bar{c})}(c)\mathbb{I}$ is minimal. Towards the top layers these have a lower error than all other coarse-grainings. In the lower depths, however, the case $\be_{1}=1/3$ gives a slight improvement over these methods. The reason for this could be that the divergence-preserving solution does not obey \cref{eq:al_conditions} exactly and the other two are not exactly divergence-preserving.  Moreover, at lower depths there are more small-scale spatial features, which means that relatively large point-wise errors may arise; thus since $\beta_1=1/3$ gives a smoother field the overall error is reduced.

\section{Gradient-Preserving Coarse-Graining}
\label{sec:GP}
 \subsection{Derivation of Gradient-Preserving Coarse-Graining}\label{sec:GP_derivation}
We also wish to compute a coarse-graining on scalar quantities such that the gradient is preserved, and we follow a similar approach to that used in \Cref{sec:DP}. Applying the definitions of coarse-graining on scalar and vector quantities, we find a condition analogous to \cref{eq:DP_general}. That is, $\lrc{\cdot}^{\al'}$ is a gradient-preserving coarse-graining at $\bar{c}\in\mathcal{C}(\bar{c})$ if there exists a coarse-graining $\EXP{\cdot}^{\be'}$ such that for all fine-grid vertices $v\in\mathcal{V}$:
\algn{
 \sum_{ \lrc{c:v\in\mathcal{V}(c)}} \frac{\abs{\bar{A}_{\bar{c}}}}{\abs{A_{c}}} \be'(\bar{c},c) \Delta \B{x}_{v,c} & =  \sum_{\bar{v}\in\bar{\mathcal{V}}(\bar{c})} \al'(\bar{v},v) \Delta \bar{\B{x}}_{\bar{v},\bar{c}} \label{eq:grad_condition_general}
}
\Cref{eq:grad_condition_general} is similar to \cref{eq:DP_general}, but there is a difference in the cells and vertices over which the sums range; therefore, the choices of $\mathcal{V}(\bar{v})$ and $\mathcal{C}(\bar{c})$ may also be different. \\
\par
Let us consider the simplest possible choice for $\mathcal{V}(\bar{v})$, which is the set containing only the fine grid vertex which has the same location as the coarse-grid vertex $\bar{v}$; this means the scalar field is projected directly onto the coarse grid with no averaging. We can do this since the coarse-grid vertices are all contained in the fine grid. Explicitly, we have $\lrc{\phi}_{\bar{v}}^{\al'} = \phi_{\bar{v}}$, or equivalently $\al'(\bar{v},v) = \delta_{\bar{v},v}$. With this choice there is, for each $v\in\mathcal{V}$, at most one term on the right hand side of \cref{eq:grad_condition_general}. For the cell averaging let $\mathcal{C}(\bar{c})$ contain precisely the fine-grid cells contained in $\bar{c}$.\\
\par
\begin{figure}[H]
     \centering
        \includegraphics[width=.6\textwidth]{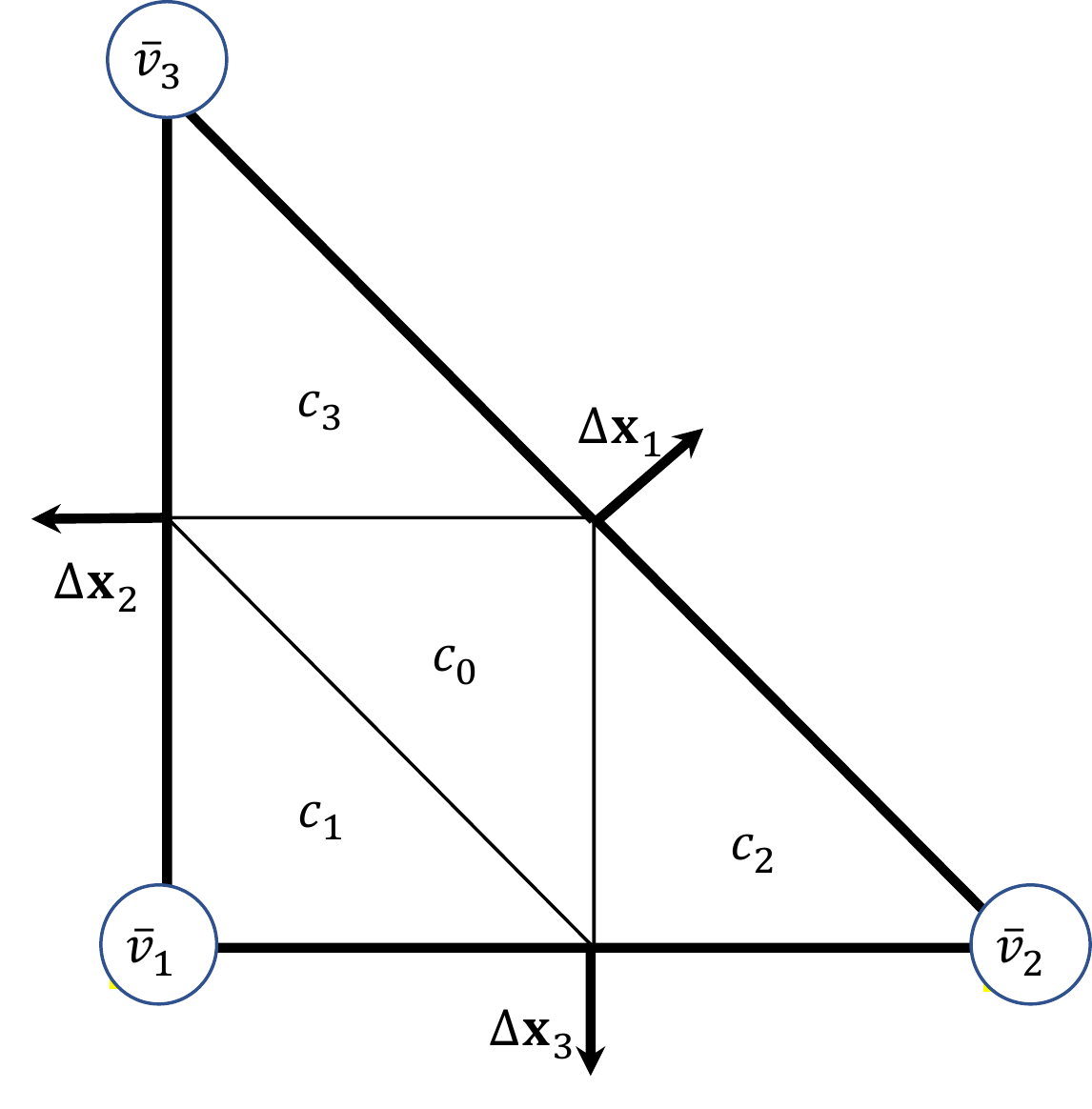}
        \captionsetup{width=0.9\linewidth}
        \caption{Averaging to preserve scalar gradients in the case $N=2$. On vertices we do direct projection. On cells we average over the four fine-grid cells contained within a coarse-grid cell.}
        \label{fig:Mesh_gradient}
\end{figure}

 
If we first consider the case $N=2$ so that, for example, if $\bar{c}$ is the cell shown in \cref{fig:Mesh_gradient}, then we have  $\mathcal{C}(\bar{c})=\lrc{c_{0},c_{1}, c_{2},c_{3}}$ and $\mathcal{V}(\bar{v}_{i}) = \lrc{\bar{v}_{i}}$ for $i=1,2,3$. Then \cref{eq:grad_condition_general} becomes, for $i=1,2,3$:
\eqsyslab{
     \beta'_{i}\Delta \B{x}_{i} & =  \Half \Delta \B{x}_{i} \\
    \be'_{0}\Delta \B{x}_{i} & = \be'_{(i+1)}\Delta \B{x}_{i-1} + \be'_{(i-1)}\Delta \B{x}_{i+1}  \label{eq:direct_proj_be0}
}{direct_proj}
 where $\beta_j := \beta(\bar{c},c_j)$ for $j=0,1,2,3$. Moreover, since the matrices $\be'_{i}$ are symmetric and positive semi-definite, their eigenvectors must be perpendicular; in particular, $\be'_{i}$ has one eigenvector $\Delta \B{x}_{i}$ so the other must be in the perpendicular direction, which we shall denote $\Delta \B{x}_{i}^{\perp}$. Writing the three vectors explicitly we have $\Delta \B{x}_1 := (1,1)\Delta x$, $\Delta \B{x}_1^\perp := (-1,1)\Delta x$, $\Delta \B{x}_2 := (-1,0)\Delta x$, $\Delta \B{x}_2^\perp := (0,-1)\Delta x$, $\Delta \B{x}_3 := (0,-1)\Delta x$, and $\Delta \B{x}_3^\perp := (-1,0)\Delta x$. Let the eigenvalue corresponding to $\Delta \B{x}_i^\perp$ be denoted $\mu_{i}\geq 0$. Then \cref{eq:direct_proj_be0} becomes:
\eqsys{
        0 & = \lrb{\be'_{0} + \mu_{2}}\Delta \B{x}_{3} + \lrb{\be'_{0} + \mu_{3}}\Delta \B{x}_{2}  \\    
        \lrb{\be'_{0} + \mu_{1} + \mu_{3} - \frac{1}{2}}\Delta \B{x}_{1}^{\perp} & =  \lrb{\be'_{0} + \mu_{3}}\Delta \B{x}_{1}\\
        \lrb{\be'_{0} + \mu_{1} + \mu_{2} - \Half}\Delta \B{x}_{1}^{\perp} & = - \lrb{\be'_{0} + \mu_{2}}\Delta \B{x}_{1}
}
This has a solution:
\algn{
\begin{split}\label{eq:solution_GP_beta}
    \be'_{0} & = \frac{1}{4}\twobytwomatrix{1-4\mu_{3}-2\mu_{1} }{2\mu_{1}-1}{2\mu_{1}-1}{1-4\mu_{2}-2\mu_{1}}  \\
    \be'_{1}  &= \frac{1}{4}\twobytwomatrix{1+2\mu_{1}}{1-2\mu_{1}}{1-2\mu_{1}}{1+2\mu_{1}} \\
     \be'_{2}   &= \frac{1}{2}\twobytwomatrix{1}{0}{0}{2\mu_{2}} \quad \quad \be'_{3}  = \frac{1}{2}\twobytwomatrix{2\mu_{3}}{0}{0}{1} 
\end{split}
}
In order that $\be'_{0}$ be positive semi-definite, we have the following restrictions on the values of $\mu_{i}$:
\eqsys{\label{eq:conditions_GP_mu}
    \mu_{1}+\mu_{2}+\mu_{3}& \leq \frac{1}{2} \\
     \lrb{\mu_{2} + \mu_{3}}\lrb{1-2\mu_{1}} &\leq 4\mu_{2}\mu_{3} 
}
Some re-arranging of \cref{eq:conditions_GP_mu} shows that the only non-negative values of $\mu_{i}$ that satisfy these conditions have $\mu_{2}=\mu_{3}=0$ and $ 0\leq \mu_{1} \leq \half$. 
However, the solution \cref{eq:solution_GP_beta} satisfies \cref{eq:be_conditionsi} if and only if $\mu_2=\mu_3=1/2$. Thus the direct projection cannot satisfy both \cref{eq:be_conditions_posdef} and \cref{eq:be_conditionsi} simultaneously. However, the fact that \cref{eq:be_conditionsi} is satisfied for the latter solution gives us that  $\bar{\nabla}\lrc{\phi}^{\al'} =\EXP{\nabla \phi}^{\be'} \approx \nabla \phi$, up to order $N^2\Delta x^2$, where $\lrc{\cdot}^{\al'}$ is the direct projection onto coarse-grid vertices. In both choices for $\EXP{\cdot}^{\be'}$ the parameter $\mu_1$ is undetermined, and so there is a family of coarse-grainings that will all give the same result when applied to a scalar gradient. We therefore have several different options for defining the coarse-graining $\EXP{\cdot}^{\be'}$, though none of these is entirely satisfactory. We shall therefore verify numerically in the next section that the coarse-graining by direct projection on scalar quantities does give a smaller error in the gradient than other methods of coarse-graining. 
  \\
\par
\begin{figure}[H]
    \centering
        \includegraphics[width=.5\textwidth]{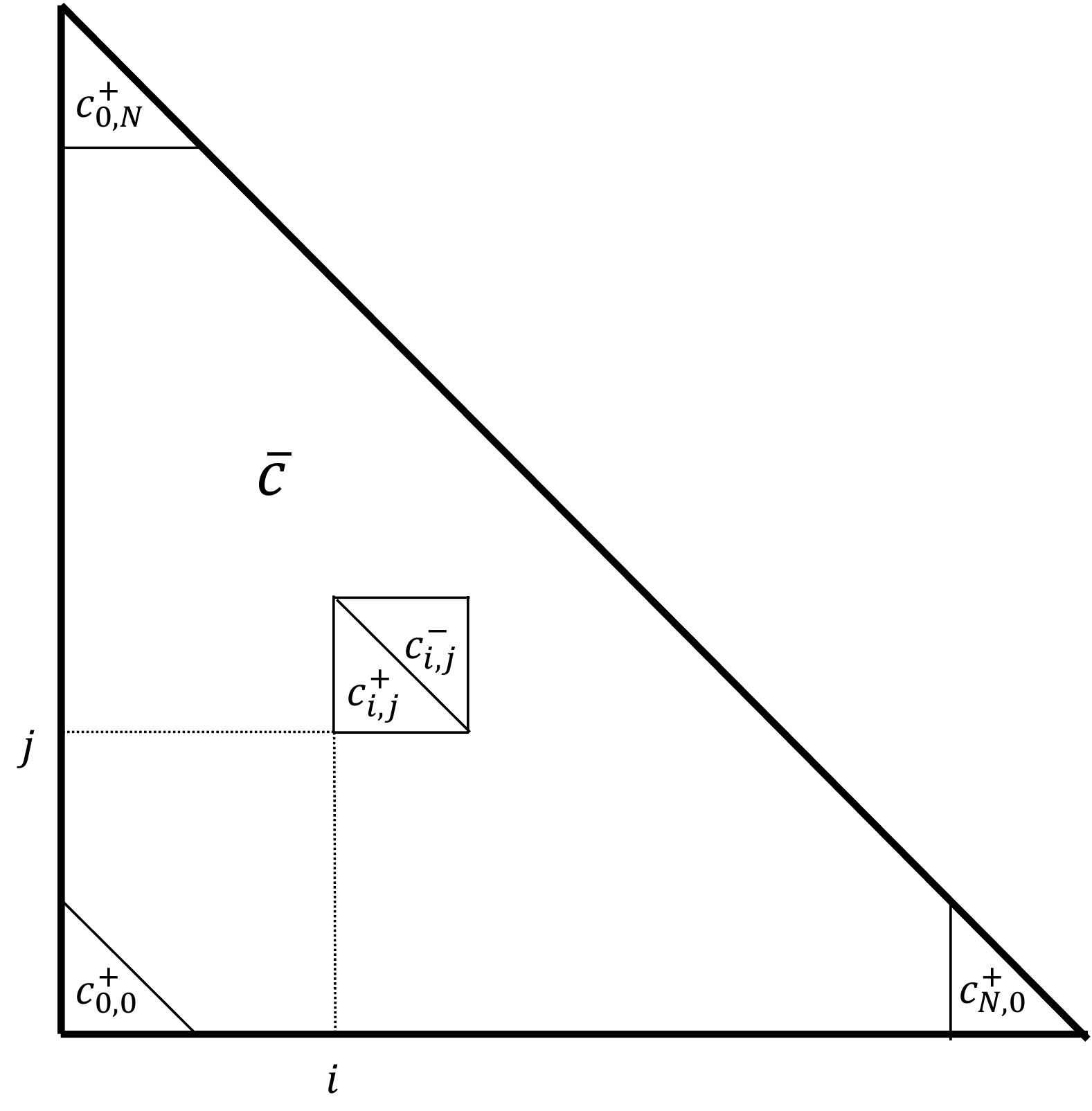}
         \captionsetup{width=0.6\linewidth}
        \caption{Fine grid cell contained within coarse grid cell for general $N$. \Cref{eq:grad_condition_general} must hold for every fine-grid vertex $v$ and every coarse-grid cell $\bar{c}$. We label the upward-pointing cells according to the position of the vertex at its right-angle, denoting as $c_{ij}^+$ the cell located at $i$ vertices horizontally and $j$ vertices vertically from the right-angle corner of $\ol{c}$. The cell denoted by $c_{ij}^-$ is the downward-pointing cell sharing a hypotenuse with $c_{ij}^+$.}
        \label{fig:GP_general}
\end{figure}

We may  generalise the results above to general $N$. Let us again suppose that $\al'(\bar{v},v) = \delta_{\bar{v},v}$ and that $\beta'\lrb{{\bar{c},c}}=0$ unless $c$ is contained in $\bar{c}$. Consider first an upward-pointing coarse-grid cell $\bar{c}$ with the fine-grid cells contained therein labelled as shown in \cref{fig:GP_general}, so that $c_{ij}^+$ is the upward pointing-cell with its right-angled vertex at position $(i, j)$ relative to the right-angled vertex of $\bar{c}$. We write $c_{ij}^-$ to denote the downward-pointing cell that shares a hypotenuse with $c_{ij}^+$. Considering condition \cref{eq:grad_condition_general} at each of the  vertices of $\bar{c}$, we find:
\eqsys{
   &\beta'\lrb{\bar{c},c_{0,0}^{+}} \Delta x_1  = \frac{1}{N}\Delta x_1 \\  &\beta'\lrb{\bar{c},c_{N-1,0}^{+}} \Delta x_2  = \frac{1}{N}\Delta x_2 \\  &\beta'\lrb{\bar{c},c_{0,N-1}^{+}} \Delta x_3  = \frac{1}{N}\Delta x_3 
   }
   At the fine-grid vertices on the boundary of the cell we find, for $i, j = 1,...,N-1$: 
   \eqsys{
&    \beta'\lrb{\bar{c},c_{i-1,0}^{-}} \Delta x_3  = \beta'\lrb{\bar{c},c_{i-1,0}^{+}}\Delta x_2 + \beta'\lrb{\bar{c},c_{i,0}^{+}}\Delta x_1  \\
   & \beta'\lrb{\bar{c},c_{0,j-1}^{-}} \Delta x_2  = \beta'\lrb{\bar{c},c_{0,j-1}^{+}}\Delta x_3 + \beta'\lrb{\bar{c},c_{0,j}^{+}}\Delta x_1\\
   & \beta'\lrb{\bar{c},c_{i-1,N-i-1}^{-}} \Delta x_1  = \beta'\lrb{\bar{c},c_{i-1,N-i}^{+}}\Delta x_2 + \beta'\lrb{\bar{c},c_{i,N-i-1}^{+}}\Delta x_3
   }
   And considering all other vertices, we have, for $i,j=1,...,N-2$ and $i+j\leq N-1$:
   \begin{subequations}
   \algn{
   \begin{split}
    0 = &\lrb{\beta'\lrb{\bar{c},c_{i,j}^{+}} - \beta'\lrb{\bar{c},c_{i-1,j-1}^{-}}}\Delta x_1 \\
    &+\lrb{\beta'\lrb{\bar{c},c_{i-1,j}^{+}} - \beta'\lrb{\bar{c},c_{i,j-1}^{-}}}\Delta x_2 \\
     &+\lrb{\beta'\lrb{\bar{c},c_{i,j-1}^{+}} - \beta'\lrb{\bar{c},c_{i-1,j}^{-}}}\Delta x_3 
    \end{split}
}
\end{subequations}
It may be verified that one possible solution to conditions is given by taking:
\eqsyslab{
      \beta'\lrb{\bar{c},c_{0,0}^+ } &= \twobytwomatrix{1/N}{0}{0}{1/N} \\
        \beta'\lrb{\bar{c},c^+_{i,0}} &= \twobytwomatrix{1/N}{0}{0}{0} \quad \text{for}\quad i>0 \\
        \beta'\lrb{\bar{c},c^+_{0,j}} &= \twobytwomatrix{0}{0}{0}{1/N} \quad \text{for}\quad j>0 
}{be_sol_N}

and $\beta'\lrb{\bar{c},c^+_{i,j}}=0$ for $ij>0$, $\beta'\lrb{\bar{c},c^-_{i,j}}=0$  for all $i,j$.  What we have therefore shown is that if we compute the gradient after a direct projection, this gives the same result as averaging the meridional (zonal) fine-grid gradients calculated on cells adjacent to the meridional (zonal) boundaries of the coarse-grid cell. \Cref{eq:be_sol_N} in the case $N=2$ corresponds to  setting $\mu_1 = \frac{1}{2}$, $\mu_2=\mu_3=0$ in \cref{eq:solution_GP_beta}. We see therefore that this method of averaging is by no means unique, and there may be others which give the same average gradient at the centre of $\bar{c}$. However, as we saw in the $N=2$ case this solution does not satisfy \cref{eq:be_conditionsi}. An alternative is to set $\beta'(\bar{c},c^{\pm}_{i,j}) = \pm\frac{1}{N}\mathbb{I}$. This satisfies \cref{eq:be_conditionsi}, though it violates \cref{eq:be_conditions_posdef}. It is clear that all of the proposed choices for $\tl{\al},\tl{\be}$ here are not integral-preserving and so are not suitable for comparing domain-averaged quantities. It remains unclear whether or not a more satisfactory solution exists, but we proceed numerically and observe in the next section that the direct projection does give a good approximation to the fine-grid gradient.

\subsection{Numerical Results}

Here we use the same model set-up as in \cref{sec:DP} to verify our results. The pressure is calculated according to the formula given in \citep{danilov_sidorenko_wang_jung_2016}, but with a linear equation of state $b(T)=\kappa(T-T_0)$ so that $p(\B{x},z) = g\lrb{\eta - \kappa \int_{z}^{0}(T- T_0) dz'}$, where $\eta$ is the sea surface height, $T$ is the temperature, and $\kappa,T_0$ are constants. We do the coarse-graining on $\eta$ and $T$, and since pressure depends linearly on these variables, this is equivalent to coarse-graining $p$ directly. If the equation of state were not linear then this would not be possible and the method for coarse-graining may vary depending on the application. This is because it is not necessarily the case that $b(\lrc{T}^{\al'}) =\lrc{b(T)}^{\al'}$. Thus if we want to preserve the buoyancy gradient we may need to search for coarse-grainings that obey $\EXP{\nabla b(T)}^{\beta'} = \bar{\nabla} b\lrb{\lrc{T}^{\al'}}$. However, this will depend on what the equation of state is and whether such a coarse-graining will exist is unclear, and beyond the scope of the present work.  \\
\par
We consider several options for the coarse graining, with the averaging done according to that shown in \cref{fig:Vv}. Explicitly, for each vertex we perform a seven-point average with weights $\al'_{0}$ on the central vertex and $\al'_{1}$ on the remaining vertices. Note that $\al'_{0}=1-6\al'_{1}$, so $0\leq \al'_{1} \leq 1/6$; therefore, we consider the set of values: $\al'_{1}=0,1/12,1/10,1/8,1/6$. The case $\al'_{1}=0$ corresponds to direct projection and should preserve the gradient, according to the results derived in \cref{sec:GP_derivation}. At the boundaries we use $\al_0'\arr \frac{\al_0'}{\al_0' +n\al_1'}$ and $\al_1'\arr \frac{\al_1'}{\al_0' +n\al_1'}$ where $n$ is the number of vertices connected by a fine-grid edge to the boundary vertex around which we average. However, in our calculations of error we exclude the boundaries so this choice will not affect our conclusions.\\
\par
\begin{figure}[H]
    \centering
    \includegraphics[width=.8\textwidth]{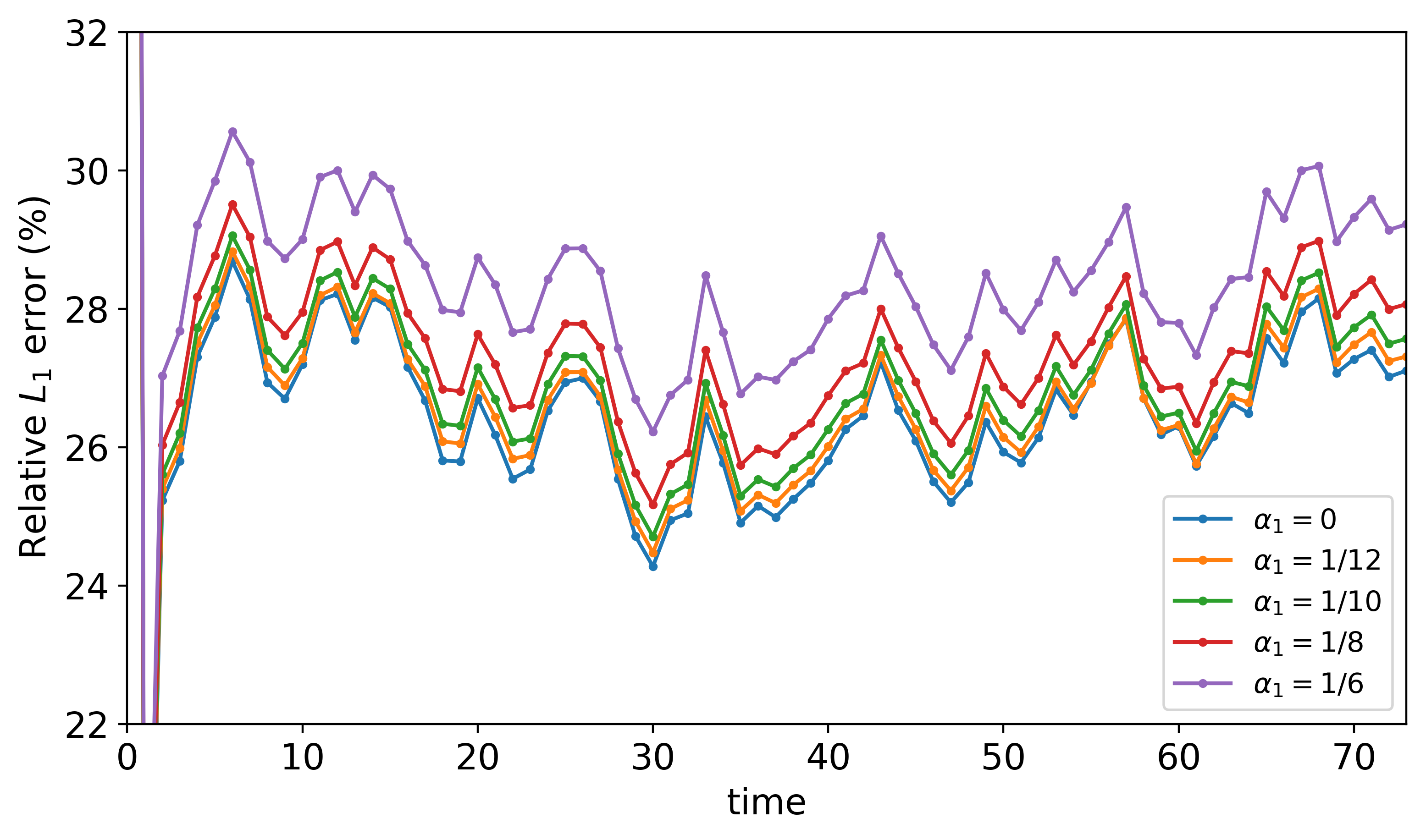}
    \caption{Relative $L_{1}$ error in $\nabla p$ at surface, compared to reference solution. The $L_{1}$ error is given by $E[\nabla p,\bar{\nabla}\bar{p}] = \frac{\int \abs{\nabla p - \bar{\nabla}\bar{p}} d^{2}x}{\int \abs{\nabla p} d^{2}x}$ and in the integral we exclude points at a perpendicular distance of $1.5^{\circ}$ from the lateral boundaries. The error is plotted for the surface layer.}
    \label{fig:grad_p_L1}
\end{figure}
\begin{figure}[H]
    \centering
    \includegraphics[width=.8\textwidth]{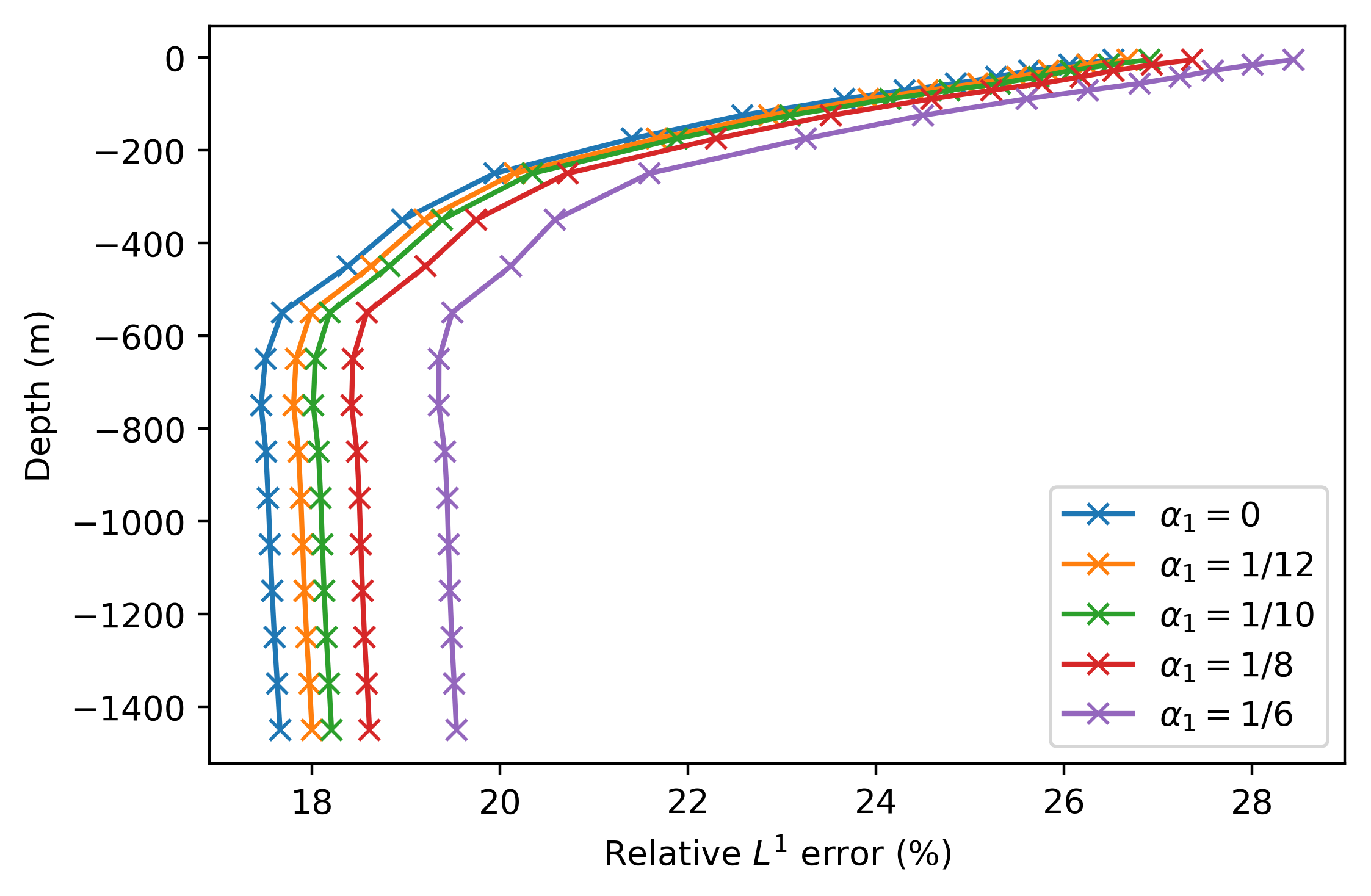}
    \caption{Time-average of relative $L^{1}$ error in pressure gradient, at different depths. Time-average excludes first five points.}
    \label{fig:dp_L1_depth}
\end{figure}

If we plot the coarse-grid gradients of the coarse-grained pressures, then we find that there is almost no visible difference between these and the fine-grid pressure gradient. We therefore do not include a plot of this. However, we plot the $L^{1}$ error of the surface pressure gradient in \cref{fig:grad_p_L1}. We see here that there is a small difference in error between the different coarse graining methods, but generally the direct projection $\al'_{1}=0$ gives the lowest error, as expected. We also see in \cref{fig:dp_L1_depth} that at other depths the direct projection continues to be the best at reducing the error and in fact the improvement becomes more noticeable at these depths.

\section{Summary and Discussion}
\label{sec:conclusion}

This study has scratched the surface of the complex and multi-faceted coarse-graining problem, and we have proposed various methods for approaching the problem; there is no objectively preferred method, but the way in which coarse-graining should be done depends on the particular application, and on what properties of the variable should be carried over to the coarse grid.  Here we have developed methods with a view to applications in Primitive Equation models and therefore have considered divergences and scalar gradients, since these are used in the calculation of important quantities such as vertical velocities and pressure gradients. We have shown that performing coarse-graining in a na\"{i}ve way has the potential to give rise to large errors. \\
\par
We have derived a general condition for divergence-preservation on B-grids and applied this to a regular triangular mesh to derive a coarse-graining which is divergence-preserving and integral-preserving away from the boundary. The solution may also be extended to boundaries which coincide in the fine and coarse grids. Where this is not the case it is possible to retain the integral-preservation property though it remains unclear how to preserve the divergence at these locations. We have further generalised the result to the case of a curved mesh and showed that a divergence-preserving coarse-graining exists which takes the curvature into account. This latter coarse-graining was the most successful in reducing errors in the vertical velocity, as we demonstrated in numerical simulations. We also showed that a divergence-preserving coarse-graining may be extended to models using $z*$-coordinates.\\
\par
In the gradient case we have derived general conditions for the averaging weights to preserve the gradient and found such weights. These weights do not have the integral-preserving property, but we show numerically that the coarse-graining consisting in a direct projection onto the coarse grid preserves the gradient better than other methods. \\
\par
This work opens a wide variety of possibilities for future development. The most obvious is to further analyse the coarse-graining problem at boundaries that do not coincide in the two grids. We could also extend the work to other operators such as the Laplacian operator, which appears in the eddy forcing as stated in \citep{mana2014toward}. Additionally, many ocean models use grids different from those considered in the present work. For example, MOM5 uses a B-grid on a square mesh \citep{GRIFFIES2000123}, while ICON \citep{KORN2017525} uses a C-grid on a triangular mesh and MITgcm \citep{marshall_adcroft_hill_perelman_heisey_1997} uses a C-grid on a square mesh. For B-grids with alternative geometries \cref{eq:DP_general} and \cref{eq:grad_condition_general} are still valid, but for grids with different variable placements new conditions would need to be derived. This work may also act as a starting point for the reverse problem, that of interpolation. i.e. defining a field on a fine grid that has the same divergence, gradient, total integral, etc as a given coarse-grid field. This could be done by using the conditions derived in the present work and ensuring that when the interpolated solution is coarse-grained back to the coarse grid, the required operator is preserved. An investigation in this direction is a promising idea for future study. 

\section*{Acknowledgements}

The author was funded by the UK Engineering and Physical Sciences Research Council (EPSRC) Centre for Doctoral Training in Mathematics of Planet Earth (MPE-CDT). He would like to thank the three anonymous reviewers for their constructive comments, as well as Pavel Berloff and Stephan Juricke for feedback on the manuscript. Also acknowledged are Darryl Holm and all the members of the Geometric Mechanics research group at Imperial College for the stimulating discussions at our Friday group meetings, as well as Colin Cotter, Peter Korn, Ruiao Hu and Pavel Perezhogin for useful discussions.

\section{Figures}

\newpage
  \bibliographystyle{elsarticle-num} 
  \bibliography{refs}

\begin{thebibliography}{10}
\expandafter\ifx\csname url\endcsname\relax
  \def\url#1{\texttt{#1}}\fi
\expandafter\ifx\csname urlprefix\endcsname\relax\def\urlprefix{URL }\fi
\expandafter\ifx\csname href\endcsname\relax
  \def\href#1#2{#2} \def\path#1{#1}\fi

\bibitem{BERLOFF2005123}
P.~S. Berloff,
  \href{http://www.sciencedirect.com/science/article/pii/S037702650400051X}{{On
  dynamically consistent eddy fluxes}}, Dynamics of Atmospheres and Oceans
  38~(3) (2005) 123 -- 146.
\newblock \href
  {http://dx.doi.org/https://doi.org/10.1016/j.dynatmoce.2004.11.003}
  {\path{doi:https://doi.org/10.1016/j.dynatmoce.2004.11.003}}.
\newline\urlprefix\url{http://www.sciencedirect.com/science/article/pii/S037702650400051X}

\bibitem{berloff2005random}
P.~S. Berloff, Random-forcing model of the mesoscale oceanic eddies, Journal of
  Fluid Mechanics 529 (2005) 71–95.
\newblock \href {http://dx.doi.org/10.1017/S0022112005003393}
  {\path{doi:10.1017/S0022112005003393}}.

\bibitem{berloff_ryzhov_shevchenko_2021}
P.~Berloff, E.~Ryzhov, I.~Shevchenko, {On dynamically unresolved oceanic
  mesoscale motions}, Journal of Fluid Mechanics 920 (2021) A41.
\newblock \href {http://dx.doi.org/10.1017/jfm.2021.477}
  {\path{doi:10.1017/jfm.2021.477}}.

\bibitem{mana2014toward}
P.~{Porta Mana}, L.~Zanna,
  \href{https://www.sciencedirect.com/science/article/pii/S1463500314000420}{{Toward
  a stochastic parameterization of ocean mesoscale eddies}}, Ocean Modelling 79
  (2014) 1--20.
\newblock \href
  {http://dx.doi.org/https://doi.org/10.1016/j.ocemod.2014.04.002}
  {\path{doi:https://doi.org/10.1016/j.ocemod.2014.04.002}}.
\newline\urlprefix\url{https://www.sciencedirect.com/science/article/pii/S1463500314000420}

\bibitem{cotter2018modelling}
C.~Cotter, D.~Crisan, D.~D. Holm, W.~Pan, I.~Shevchenko, {Modelling uncertainty
  using stochastic transport noise in a 2-layer quasi-geostrophic model}
  (2020).
\newblock \href {http://arxiv.org/abs/1802.05711} {\path{arXiv:1802.05711}}.

\bibitem{cotter2019numerically}
C.~Cotter, D.~Crisan, D.~D. Holm, W.~Pan, I.~Shevchenko,
  \href{https://doi.org/10.1137/18M1167929}{{Numerically Modeling Stochastic
  Lie Transport in Fluid Dynamics}}, Multiscale Modeling \& Simulation 17~(1)
  (2019) 192--232.
\newblock \href {http://arxiv.org/abs/https://doi.org/10.1137/18M1167929}
  {\path{arXiv:https://doi.org/10.1137/18M1167929}}, \href
  {http://dx.doi.org/10.1137/18M1167929} {\path{doi:10.1137/18M1167929}}.
\newline\urlprefix\url{https://doi.org/10.1137/18M1167929}

\bibitem{holm2015}
D.~D. Holm,
  \href{https://royalsocietypublishing.org/doi/abs/10.1098/rspa.2014.0963}{{Variational
  principles for stochastic fluid dynamics}}, Proceedings of the Royal Society
  A: Mathematical, Physical and Engineering Sciences 471~(2176) (2015)
  20140963.
\newblock \href
  {http://arxiv.org/abs/https://royalsocietypublishing.org/doi/pdf/10.1098/rspa.2014.0963}
  {\path{arXiv:https://royalsocietypublishing.org/doi/pdf/10.1098/rspa.2014.0963}},
  \href {http://dx.doi.org/10.1098/rspa.2014.0963}
  {\path{doi:10.1098/rspa.2014.0963}}.
\newline\urlprefix\url{https://royalsocietypublishing.org/doi/abs/10.1098/rspa.2014.0963}

\bibitem{VICHI200789}
M.~Vichi, N.~Pinardi, S.~Masina,
  \href{https://www.sciencedirect.com/science/article/pii/S0924796306001084}{{A
  generalized model of pelagic biogeochemistry for the global ocean ecosystem.
  Part I: Theory}}, {Journal of Marine Systems} 64~(1) (2007) 89--109,
  contributions from Advances in Marine Ecosystem Modelling Research, 27-29
  June, 2005, Plymouth, UK.
\newblock \href
  {http://dx.doi.org/https://doi.org/10.1016/j.jmarsys.2006.03.006}
  {\path{doi:https://doi.org/10.1016/j.jmarsys.2006.03.006}}.
\newline\urlprefix\url{https://www.sciencedirect.com/science/article/pii/S0924796306001084}

\bibitem{ENGLAND_MAIERREIMER2001}
{England, Matthew H. and Maier-Reimer, Ernst},
  \href{https://agupubs.onlinelibrary.wiley.com/doi/abs/10.1029/1998RG000043}{{Using
  chemical tracers to assess ocean models}}, {Reviews of Geophysics} 39~(1)
  (2001) 29--70.
\newblock \href
  {http://arxiv.org/abs/https://agupubs.onlinelibrary.wiley.com/doi/pdf/10.1029/1998RG000043}
  {\path{arXiv:https://agupubs.onlinelibrary.wiley.com/doi/pdf/10.1029/1998RG000043}},
  \href {http://dx.doi.org/https://doi.org/10.1029/1998RG000043}
  {\path{doi:https://doi.org/10.1029/1998RG000043}}.
\newline\urlprefix\url{https://agupubs.onlinelibrary.wiley.com/doi/abs/10.1029/1998RG000043}

\bibitem{England_Rahmstorf}
M.~H. England, S.~Rahmstorf,
  \href{https://journals.ametsoc.org/view/journals/phoc/29/11/1520-0485_1999_029_2802_sovrar_2.0.co_2.xml}{{Sensitivity
  of Ventilation Rates and Radiocarbon Uptake to Subgrid-Scale Mixing in Ocean
  Models}}, Journal of Physical Oceanography 29~(11) (1999) 2802 -- 2828.
\newblock \href
  {http://dx.doi.org/10.1175/1520-0485(1999)029<2802:SOVRAR>2.0.CO;2}
  {\path{doi:10.1175/1520-0485(1999)029<2802:SOVRAR>2.0.CO;2}}.
\newline\urlprefix\url{https://journals.ametsoc.org/view/journals/phoc/29/11/1520-0485_1999_029_2802_sovrar_2.0.co_2.xml}

\bibitem{CABOS}
W.~Cabos, A.~de~la Vara, F.~J. {\'A}lvarez-Garc{\'\i}a, E.~S{\'a}nchez,
  K.~Sieck, J.-I. P{\'e}rez-Sanz, N.~Limareva, D.~V. Sein,
  \href{https://doi.org/10.1007/s00382-020-05238-x}{{Impact of ocean-atmosphere
  coupling on regional climate: the Iberian Peninsula case}}, Climate Dynamics
  54~(9) (2020) 4441--4467.
\newblock \href {http://dx.doi.org/10.1007/s00382-020-05238-x}
  {\path{doi:10.1007/s00382-020-05238-x}}.
\newline\urlprefix\url{https://doi.org/10.1007/s00382-020-05238-x}

\bibitem{Xue2019}
P.~Xue, P.~Malanotte-Rizzoli, J.~Wei, E.~A.~B. Eltahir,
  \href{https://agupubs.onlinelibrary.wiley.com/doi/abs/10.1029/2019JC014978}{Coupled
  ocean-atmosphere modeling over the maritime continent: A review}, Journal of
  Geophysical Research: Oceans 125~(6) (2020) e2019JC014978, e2019JC014978
  2019JC014978.
\newblock \href
  {http://arxiv.org/abs/https://agupubs.onlinelibrary.wiley.com/doi/pdf/10.1029/2019JC014978}
  {\path{arXiv:https://agupubs.onlinelibrary.wiley.com/doi/pdf/10.1029/2019JC014978}},
  \href {http://dx.doi.org/https://doi.org/10.1029/2019JC014978}
  {\path{doi:https://doi.org/10.1029/2019JC014978}}.
\newline\urlprefix\url{https://agupubs.onlinelibrary.wiley.com/doi/abs/10.1029/2019JC014978}

\bibitem{OASIS}
A.~Craig, S.~Valcke, L.~Coquart,
  \href{https://gmd.copernicus.org/articles/10/3297/2017/}{{Development and
  performance of a new version of the OASIS coupler, OASIS3-MCT\_3.0}},
  Geoscientific Model Development 10~(9) (2017) 3297--3308.
\newblock \href {http://dx.doi.org/10.5194/gmd-10-3297-2017}
  {\path{doi:10.5194/gmd-10-3297-2017}}.
\newline\urlprefix\url{https://gmd.copernicus.org/articles/10/3297/2017/}

\bibitem{CMIP2016}
V.~Eyring, S.~Bony, G.~A. Meehl, C.~A. Senior, B.~Stevens, R.~J. Stouffer,
  K.~E. Taylor,
  \href{https://gmd.copernicus.org/articles/9/1937/2016/}{Overview of the
  coupled model intercomparison project phase 6 (cmip6) experimental design and
  organization}, Geoscientific Model Development 9~(5) (2016) 1937--1958.
\newblock \href {http://dx.doi.org/10.5194/gmd-9-1937-2016}
  {\path{doi:10.5194/gmd-9-1937-2016}}.
\newline\urlprefix\url{https://gmd.copernicus.org/articles/9/1937/2016/}

\bibitem{vallis2017atmospheric}
G.~K. Vallis, {Atmospheric and Oceanic Fluid Dynamics: Fundamentals and
  Large-Scale Circulation}, 2nd Edition, Cambridge University Press, 2017.
\newblock \href {http://dx.doi.org/10.1017/9781107588417}
  {\path{doi:10.1017/9781107588417}}.

\bibitem{danilov_sidorenko_wang_jung_2016}
S.~Danilov, D.~Sidorenko, Q.~Wang, T.~Jung, {The Finite-volumE Sea ice-Ocean
  Model} ({FESOM2}), Geoscientific Model Development Discussions (2016)
  1--44\href {http://dx.doi.org/10.5194/gmd-2016-260}
  {\path{doi:10.5194/gmd-2016-260}}.

\bibitem{ARAKAWA_1977}
A.~Arakawa, V.~R. Lamb,
  \href{https://www.sciencedirect.com/science/article/pii/B9780124608177500094}{{Computational
  Design of the Basic Dynamical Processes of the UCLA General Circulation
  Model}}, Methods in Computational Physics: Advances in Research and
  Applications 17 (1977) 173--265.
\newblock \href
  {http://dx.doi.org/https://doi.org/10.1016/B978-0-12-460817-7.50009-4}
  {\path{doi:https://doi.org/10.1016/B978-0-12-460817-7.50009-4}}.
\newline\urlprefix\url{https://www.sciencedirect.com/science/article/pii/B9780124608177500094}

\bibitem{barth2018finite}
T.~Barth, R.~Herbin, M.~Ohlberger,
  \href{https://onlinelibrary.wiley.com/doi/abs/10.1002/9781119176817.ecm2010}{{Finite
  Volume Methods: Foundation and Analysis}}, American Cancer Society, 2017, pp.
  1--60.
\newblock \href
  {http://arxiv.org/abs/https://onlinelibrary.wiley.com/doi/pdf/10.1002/9781119176817.ecm2010}
  {\path{arXiv:https://onlinelibrary.wiley.com/doi/pdf/10.1002/9781119176817.ecm2010}},
  \href {http://dx.doi.org/https://doi.org/10.1002/9781119176817.ecm2010}
  {\path{doi:https://doi.org/10.1002/9781119176817.ecm2010}}.
\newline\urlprefix\url{https://onlinelibrary.wiley.com/doi/abs/10.1002/9781119176817.ecm2010}

\bibitem{DANILOV201214}
{S. Danilov},
  \href{https://www.sciencedirect.com/science/article/pii/S1463500312000194}{{Two
  finite-volume unstructured mesh models for large-scale ocean modeling}},
  {Ocean Modelling} 47 (2012) 14--25.
\newblock \href
  {http://dx.doi.org/https://doi.org/10.1016/j.ocemod.2012.01.004}
  {\path{doi:https://doi.org/10.1016/j.ocemod.2012.01.004}}.
\newline\urlprefix\url{https://www.sciencedirect.com/science/article/pii/S1463500312000194}

\bibitem{juricke_danilov_kutsenko_oliver_2019}
S.~Juricke, S.~Danilov, A.~Kutsenko, M.~Oliver,
  \href{https://www.sciencedirect.com/science/article/pii/S1463500318303846}{{Ocean
  kinetic energy backscatter parametrizations on unstructured grids: Impact on
  mesoscale turbulence in a channel}}, Ocean Modelling 138 (2019) 51--67.
\newblock \href
  {http://dx.doi.org/https://doi.org/10.1016/j.ocemod.2019.03.009}
  {\path{doi:https://doi.org/10.1016/j.ocemod.2019.03.009}}.
\newline\urlprefix\url{https://www.sciencedirect.com/science/article/pii/S1463500318303846}

\bibitem{GRIFFIES2000123}
S.~M. Griffies, C.~B\"{o}ning, F.~O. Bryan, E.~P. Chassignet, R.~Gerdes,
  H.~Hasumi, A.~Hirst, A.-M. Treguier, D.~Webb,
  \href{https://www.sciencedirect.com/science/article/pii/S1463500300000147}{{Developments
  in ocean climate modelling}}, {Ocean Modelling} 2~(3) (2000) 123--192.
\newblock \href
  {http://dx.doi.org/https://doi.org/10.1016/S1463-5003(00)00014-7}
  {\path{doi:https://doi.org/10.1016/S1463-5003(00)00014-7}}.
\newline\urlprefix\url{https://www.sciencedirect.com/science/article/pii/S1463500300000147}

\bibitem{KORN2017525}
P.~Korn,
  \href{https://www.sciencedirect.com/science/article/pii/S0021999117301961}{{Formulation
  of an unstructured grid model for global ocean dynamics}}, {Journal of
  Computational Physics} 339 (2017) 525--552.
\newblock \href {http://dx.doi.org/https://doi.org/10.1016/j.jcp.2017.03.009}
  {\path{doi:https://doi.org/10.1016/j.jcp.2017.03.009}}.
\newline\urlprefix\url{https://www.sciencedirect.com/science/article/pii/S0021999117301961}

\bibitem{marshall_adcroft_hill_perelman_heisey_1997}
J.~Marshall, A.~Adcroft, C.~Hill, L.~Perelman, C.~Heisey, {A finite-volume,
  incompressible Navier Stokes model for studies of the ocean on parallel
  computers}, Journal of Geophysical Research: Oceans 102~(C3) (1997)
  5753--5766.
\newblock \href {http://dx.doi.org/10.1029/96jc02775}
  {\path{doi:10.1029/96jc02775}}.

\end{thebibliography}






\end{document}